\documentclass[english,11pt, journal,draft,onecolumn]{IEEEtran}
\usepackage[T1]{fontenc}
\usepackage[latin9]{inputenc}
\usepackage{stmaryrd}
\usepackage{amssymb}
\usepackage{mathrsfs}
\usepackage{amsfonts}  
\usepackage[dvips,final]{graphicx}
\usepackage[dvips]{color}
\usepackage{epsfig,amssymb,amsmath,amsthm}
\usepackage{graphicx}
\usepackage{subfigure}

\usepackage{booktabs}

\usepackage{dsfont}
\usepackage{esint}

\usepackage{relsize}\usepackage{subdepth}\allowdisplaybreaks

\newcommand{\beq}{\begin{equation}}
\newcommand{\eeq}{\end{equation}}
\newcommand{\beqn}{\begin{eqnarray}}
\newcommand{\eeqn}{\end{eqnarray}}

\usepackage{babel}
  \theoremstyle{definition}
  \newtheorem{defn}{\protect\definitionname}
  \theoremstyle{plain}
  \newtheorem{thm}{\protect\theoremname}
  
  \theoremstyle{plain}
  
\usepackage{babel}
\providecommand{\definitionname}{Definition}

\providecommand{\theoremname}{Theorem}

\begin{document}

\title{MIMO-MC Radar: A MIMO Radar Approach Based on Matrix Completion}

\author{\IEEEauthorblockN{Shunqiao Sun, Waheed U. Bajwa and Athina P. Petropulu\\}
\IEEEauthorblockA{Department of Electrical \& Computer Engineering \\ Rutgers, the State
University of New Jersey\\
Email: \{shunq.sun, waheed.bajwa, athinap\}@rutgers.edu
}
\thanks{This work was supported by the Office of Naval Research under Grant
ONR-N-00014-12-1-0036.}}

\maketitle

\begin{abstract}
In a typical MIMO radar scenario,  transmit nodes transmit  orthogonal waveforms, while
 each receive node performs matched filtering with the known set of transmit waveforms, and forwards  the results to the fusion center. Based on the data it receives from multiple antennas, the  fusion center  formulates a matrix,
 which, in
conjunction with standard array processing schemes, such as MUSIC, leads to target detection and parameter estimation.
In MIMO radars with compressive sensing (MIMO-CS), the data matrix is formulated by each receive node forwarding   a small number of compressively obtained samples.
In this paper, it is shown  that
under certain conditions, in  both sampling cases, the data matrix at the fusion center is low-rank, and thus can be recovered based on knowledge of a small subset of its entries via matrix completion (MC) techniques.
Leveraging the low-rank property of that matrix,  we propose a new MIMO radar approach, termed, MIMO-MC radar,
in which
each receive node either performs matched filtering with a small number of randomly
selected dictionary waveforms or obtains sub-Nyquist samples of the received signal at random sampling instants,  and forwards the results to a fusion center. Based on the received samples, and with knowledge of the sampling scheme,
 the fusion center partially fills the data matrix and subsequently
  applies MC techniques to estimate the full matrix. MIMO-MC radars share the advantages of the recently proposed MIMO-CS radars, i.e., high resolution with reduced amounts of data, but unlike MIMO-CS radars do not require grid discretization.
  The MIMO-MC radar concept is illustrated through a linear uniform array configuration, and its target estimation performance is demonstrated via simulations.

\end{abstract}
\section{Introduction}
\label{sec:intro} Multiple-input and multiple-output (MIMO) radar systems
have received considerable attention in recent years due to their superior
resolution  \cite{Haimovich,Stoica: 07m,chen}. The MIMO radars using compressed sensing (MIMO-CS) maintain the MIMO radars advantages, while  significantly reducing
the required measurements per receive antenna \cite{Herman,yao}. In
MIMO-CS radars, the target parameters are estimated by exploiting the
sparsity of targets in the angle, Doppler and range space, referred to as the
\emph{target space}; the target space
is discretized into a fine grid, based on which a compressive sensing
matrix is constructed, and the target is estimated via sparse signal
recovery techniques, such as the Dantzig selector \cite{yao}.
 However, the performance of CS-based MIMO radars degrades
when  targets fall between grid points, a case also known as basis mismatch
\cite{chi,bajwa}.

In this paper, a novel approach to lower-complexity, higher-resolution radar is proposed, termed  MIMO-MC radars, which stands for MIMO radars using matrix completion (MC). MIMO-MC radars achieve the advantages of MIMO-CS radars without requiring grid discretization.
Matrix completion is of interest in cases in which  we are constrained to observe only a subset of the entries of an $n_1\times n_2$ matrix, because the cost of collecting all entries of a high dimensional matrix is high.
 If a matrix is low rank and satisfies certain conditions \cite{Candes&Recht2009}, it can be recovered \emph{exactly} based on observations of a small number of
 its randomly selected entries.
 There are several MC techniques in the literature
 \cite{Candes&Recht2009,completion1,completion2,Keshavan,Dai,Vandereycken}.
 For example, in \cite{Candes&Recht2009,completion1,completion2}, recovery can be  performed by
 solving a nuclear norm optimization problem, which basically finds the matrix with the smallest
 nuclear norm out of all possible matrices that fit the observed entries.  Other matrix completion techniques
  are based on non-convex optimization using matrix manifolds, such as Grassmann
 manifold \cite{Keshavan,Dai}, and Riemann manifolds \cite{Vandereycken}.

In a typical MIMO radar scenario \cite{chen}, transmit nodes transmit  orthogonal waveforms, while
 each receive node performs matched filtering with the known set of transmit waveforms, and forwards  the results to the fusion center. Based on the data it receives from multiple antennas, the  fusion center  formulates a matrix, which, in
conjunction with standard array processing schemes, such as MUSIC \cite{trees}, leads to target detection and estimation.
In MIMO-CS radars,  each receive nodes  uses a compressive receiver  to obtain a small number of samples, which are then
 forwarded to the fusion center \cite{Herman}\cite{yao}. Again, the fusion center can formulate a matrix based on the data forwarded by all receive nodes, which
 is then used for target estimation.
In the latter case, since no matched filtering is performed,  the waveforms do not need to be orthogonal.
In this paper, we show  that
under certain conditions, in both aforementioned sampling  cases, the data matrix at the fusion center is low-rank, which means that it can be recovered based on knowledge of a small subset of its entries via matrix completion (MC) techniques.
Leveraging the low-rank property of that matrix, we propose MIMO-MC radar, in which,
each receive antenna  either performs matched filtering with a small number of dictionary waveforms or
obtains sub-Nyquist samples of the received signal and forwards the results to a fusion center. Based on the samples forwarded by all receive nodes, and with knowledge of the sampling scheme,
 the fusion center applies MC to estimate the full matrix.
Although the proposed ideas apply to arbitrary transmit and receive array configurations, in which the antennas are not physically connected, in this paper we illustrate the idea through a linear uniform array configuration.
The properties and performance of the proposed scheme are demonstrated via simulations.
 Compared to  MIMO-CS radars, MIMO-MC radars have the same
advantage in terms of reduction of samples needed for accurate estimation, while they
 avoid the basis mismatch issue, which is inherent in MIMO-CS radar systems. Preliminary results of this work have been published in \cite{Sun_icassp}.

{\it Relation to prior work} - Array signal processing with matrix
completion has been studied in \cite{Waters,wangxin}. To the best of
our knowledge,  matrix completion has not been exploited for target
estimation in colocated MIMO radar. Our paper is related to the ideas in
\cite{wangxin} in the sense that matrix completion is applied to the received
data matrix formed by an array. However, due to the unique structure of the
received signal in MIMO radar, the problem formulation and treatment in here
is different than that in \cite{wangxin}. 

The  paper is organized as follows.
Background on noisy
matrix completion and colocated MIMO radars
 is provided  in Section \ref{model}. The proposed MIMO-MC radar approach  is presented in
Section \ref{mc_radar}.
Simulations results are given  in Section \ref{numerical}. Finally, Section \ref{conclusion} provides some concluding remarks.

 \emph{Notation}: Lower-case and upper-case letters in bold denote
 vectors and matrices, respectively. Superscripts
 $ (\cdot)^{H}$ and $(\cdot)^{T}$ denote
Hermitian transpose and  transpose, respectively.
${\mathbf 0}_{L\times M}$
and ${\mathbf 1}_{L\times M}$  denote an $L\times M$
matrix with all ``$0$'' and all ``$1$'' entries, respectively. ${\mathbf I}_{ M}$ represents an identity matrix of size $M$.  $\otimes$ denotes the Kronecker tensor product. ${\left\| {\bf{X}} \right\|_*}$ is the nuclear norm, i.e.,
 sum of the singular values; ${\left\| {\bf{X}} \right\|}$ is the operator norm; ${\left\| {\bf{X}} \right\|_F}$ is the Frobenius norm; ${\bf X}^*$ denotes the adjoint of $\bf X$.

\section{Preliminaries }
\label{model}
\subsection{Matrix Completion}
\label{matrix_completion}

In this section we provide a brief overview of the problem of recovering a rank $r$
matrix ${\bf M} \in {{\mathbb{C}}^{{n_1} \times {n_2}}}$ based on partial
knowledge of its entries using the method of \cite{Candes&Recht2009}\cite{completion1}\cite{completion2}.

Let us define the observation operation  ${\bf{Y}} = {{\mathcal P}_\Omega }\left( {\bf{M}} \right)$ as
\begin{align}
{\left[{\bf{Y}}\right]_{ij}} = \left\{ \begin{array}{l}
 {\left[{\bf{M}}\right]_{ij}},\;\left( {i,j} \right) \in \Omega  \\
 0,\quad\; {\rm{ otherwise}} \\
 \end{array} \right.
 \end{align}
where $\Omega$ is the set of indices of observed entries with cardinality $m$.
According to \cite{completion1}, when $\bf M$ is low-rank and meets  certain conditions  (see (\textbf{A0}) and (\textbf{A1}), later in this section),  $\bf M$ can be estimated by solving a  nuclear norm optimization problem
\begin{align} \label{exact_mc}
 &\min \;{\left\| {\bf{X}} \right\|_*} \nonumber \\
 &{\rm{s}}{\rm{.t}}{\rm{. }} \;\; {{\mathcal P}_\Omega }\left( {\bf{X}} \right) = {{\mathcal P}_\Omega }\left( {\bf{M}} \right)
 \end{align}
where ${\left\|  \cdot  \right\|_*}$ denotes the nuclear norm, i.e., the sum of singular values of $\bf X$.

In practice, the observations are typically  corrupted by noise,  i.e., ${\left[
{\bf{Y}} \right]_{ij}} = {\left[ {\bf{M}} \right]_{ij}} + {\left[ {\bf{E}}
\right]_{ij}}, \left( {i,j} \right) \in \Omega$, where, ${\left[ {\bf{E}}
\right]_{ij}}$ represents noise.
In that case, it holds that ${{\mathcal P}_\Omega }\left( {\bf{Y}} \right)
= {{\mathcal P}_\Omega }\left( {\bf{M}} \right) + {{\mathcal P}_\Omega
}\left( {\bf{E}} \right)$, and the
 completion of $\bf M$ is done by solving the following optimization problem \cite{completion2}
\begin{align} \label{robust_mc}
 \min \;\;&{\left\| {\bf{X}} \right\|_*} \quad  \quad \nonumber\\
 {\rm{s.t.}}\;\; &{\left\| {{{\mathcal P}_\Omega }\left( {{\bf{X - Y}}} \right)} \right\|_F} \le \delta .
 \end{align}
Assuming that the noise is  zero-mean, white,  $\delta>0$ is a parameter related to the noise variance,  $\sigma^2$, as $\delta^2 = (m+\sqrt{8m})\sigma^2$ \cite{Candes&Recht2009}.

The conditions for successful matrix completion involve the notion of incoherence, which is defined next
\cite{Candes&Recht2009}.

\begin{defn}
Let $U$ be a subspace of ${{\bf{\mathbb C}}^{{n_1}}}$ of dimension $r$ that is spanned by the set of orthogonal vectors ${\left\{ {{{\bf{u}}_i} \in {{\bf{\mathbb C}}^{{n_1}}}} \right\}_{i = 1,\ldots,r}}$, $P_U$ be the orthogonal projection onto $U$, i.e.,  ${P_U} = \sum\limits_{1 \le i \le
r} {{{\bf{u}}_i}{\bf{u}}_i^H}$, and  ${\bf e}_i$ be the standard basis vector whose $i$th element is $1$. The coherence of $U$ is defined as
\begin{align}
\mu \left( U \right) = \frac{{{n_1}}}{r}\mathop {\max }\limits_{1 \le i \le {n_1}}
{\left\| {{P_U}{{\bf{e}}_i}} \right\|^2} \quad \in \left[ {1,\frac{n_1}{r}} \right] .
\end{align}
\end{defn}

Let the compact singular value decomposition (SVD) of $\bf
M$ be  ${\bf{M}} = \sum\limits_{k = 1}^r {{\rho
_k}{{\bf{u}}_k}{\bf{v}}_k^H}$, where $\rho_k, k=1,\ldots,r$ are the singular values, and ${\bf u}_k$ and ${\bf v}_k$ the corresponding left and right singular vectors, respectively. Let $U,V$ be the subspaces spanned by ${\bf u}_k$ and ${\bf v}_k$, respectively.
Matrix ${\bf M}$ has coherence with parameters $\mu_0$ and $\mu_1$ if \\
({\bf{A0}}) $\max \left( {\mu \left( U \right),\mu \left( V \right)} \right) \le {\mu _0}$ for some positive $\mu_0$.\\
({\bf{A1}}) The maximum element  of the
$n_1\times n_2$ matrix $\sum\limits_{1 \le i \le r} {{{\bf{u}}_i}{\bf{v}}_i^H}$ is bounded by ${\mu _1}\sqrt {{r \mathord{\left/
 {\vphantom {r {\left( {{n_1}{n_2}} \right)}}} \right.
 \kern-\nulldelimiterspace} {\left( {{n_1}{n_2}} \right)}}} $ in absolute value, for some positive $\mu_1$.\\
In fact, it was shown in \cite{Candes&Recht2009} that if ({\bf{A0}}) holds, then   ({\bf{A1}}) also holds with ${\mu _1} \le {\mu _0}\sqrt r $.

Now, suppose that matrix $ {\bf M} \in {\mathbb C}^{n_1 \times n_2}$
satisfies ($\mathbf{A0}$) and ($\mathbf{A1}$).
 The following lemma gives a probabilistic bound for the number of entries, $m$, needed to estimate  ${\bf M}$.


\begin{thm} \label{sample_lemma} \cite{Candes&Recht2009}
Suppose that we observe
$m$ entries of the rank$-r$ matrix $M \in {\mathbb C}^{n_1 \times n_2}$, with matrix coordinates sampled uniformly
at random. Let $n= \max \{n_1,n _2\}$. There exist constants $C$ and $c$ such that if
\begin{equation}
m\geq C\max\left\{ \mu_{1}^{2},\mu_{0}^{1/2}\mu_{1},\mu_{0}n^{1/4}\right\} nr\beta\log n  \quad   \nonumber
\end{equation}
for some $\beta>2$, the minimizer to the program of (\ref{exact_mc})
is unique and equal to $\mathbf{M}$ with probability at least $1-cn^{-\beta}$.

For $r\leq\mu_{0}^{-1}n^{1/5}$ the bound can be improved to
\begin{equation}
m\geq C\mu_{0}n^{6/5}r\beta\log n,  \nonumber
\end{equation}
without affecting the probability of success.
\end{thm}

Theorem \ref{sample_lemma} implies that the lower the coherence parameter $\mu_0$, the fewer entries of ${\bf M}$ are required to estimate ${\bf M}$. The smallest possible value for $\mu_0$ is $1$.

Further, \cite{completion2} establishes that, when observations are
corrupted with white zero-mean Gaussian noise with variance ${\sigma^2}$, when solving (\ref{robust_mc}),
the recovery error is bounded as
\begin{equation}
{\left\| {{\bf{M - \hat M}}} \right\|_F} \le 4\sqrt {\frac{1}{p}{\left( {2 +
p} \right)\min \left( {{n_1},{n_2}} \right)}} \delta + 2\delta,
\end{equation}
where $p = \frac{m}{{{n_1}{n_2}}}$ is the fraction of observed entries, and $\delta^2 = (m+\sqrt{8m})\sigma^2$.

\subsection{Colocated MIMO Radars }

\begin{figure}
\centering
\includegraphics[width=5.3in]{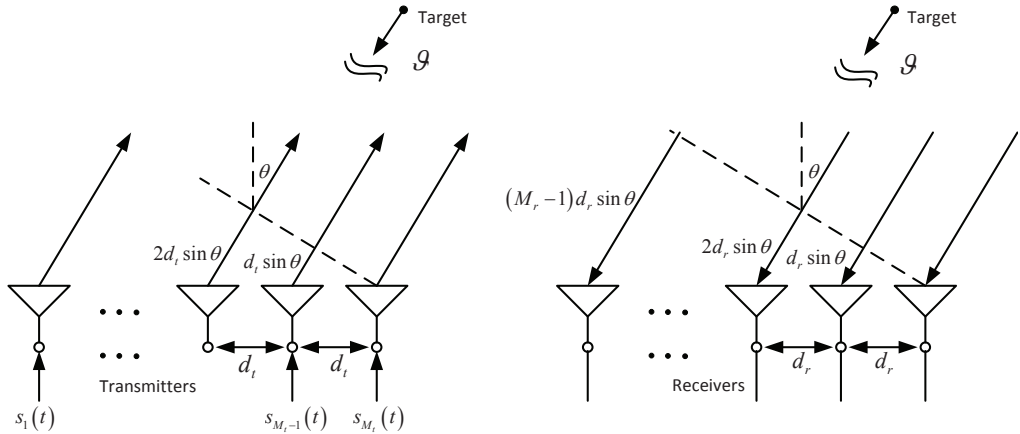}
\caption{Colocated MIMO radar system under ULA model. There are $M_t$ transmit antennas and $M_r$ receive antennas. The target is in direction $\theta$ and moving with speed $\vartheta$. }\label{system}  
\end{figure}

Let us  consider a MIMO pulse radar system that employs colocated transmit and receive
antennas, as shown in Fig. \ref{system}. We use $M_t$ and $M_r$ to denote
the numbers of transmit and receive antennas, respectively. Although
our results can be extended to an arbitrary antenna configuration, we
illustrate the ideas for uniform linear arrays (ULAs). The inter-element spacing in the transmit and receive arrays is denoted by  $d_t$ and $d_r$, respectively.
The pulse duration is $T_p$, and the pulse repetition interval is $T_{PRI}$. The
 waveform of the $i$th transmit antenna is ${s_i}\left( \tau \right) = \sqrt {\frac{E}{{{M_t}}}} {\phi _i}\left( \tau \right)
 $,
where $E$ is the total energy for all the transmit antennas, and ${\phi _i}\left( \tau \right), \ i=1,\ldots,M_t$ are orthonormal. The waveforms are transmitted over a carrier with wavelength $\lambda$.
Let us consider a scenario with $K$ point targets in the far field at angles $\theta_k, k=1,\ldots,K$, each moving with speed $\vartheta_k$.

The following assumptions are made:
\begin{itemize}
\item The transmit waveforms are narrowband, i.e., $\frac{1}{T_p}\ll \frac{c}{\lambda}$, where $c$ is the speed of light.
\item The target reflection
coefficients $\left\{ {{\beta _k}} \right\},k = 1,\ldots,K$  are complex and
remain constant during a number of  pulses, $Q$. Also, all parameters related to the array configuration
remain constant during the $Q$ pulses.
\item The delay spread in the receive signals is smaller than the temporal support of pulse $T_p$.
\item The Doppler spread of the receive signals is much smaller than the bandwidth of the pulse, i.e.,  $\frac{{2 \vartheta }}{\lambda } \ll \frac{1}{T_p}$.
\end{itemize}

Under the narrowband transmit waveform assumption, the delay spread in the baseband signals can be ignored. For slowly moving targets, the Doppler shift within a pulse can be ignored, while the Doppler  changes from pulse to pulse. Thus, if we express time as $t=q{T_{PRI}} + \tau$, where   $q$ is the pulse index (or slow time) and $\tau \in [0,T_p]$ is the time within a pulse (or fast time), the Doppler shift will depend on $q$ only, and   the received signal at the $l$-th receive antenna can be approximated as \cite{chen}
\begin{align}
{x_l}\left( {q{T_{PRI}} + \tau  + \frac{{2d}}{c}} \right) &\approx \sum\limits_{k = 1}^K {{\beta _k}{e^{j\frac{{2\pi }}{\lambda }\left( {2{\vartheta _k}\left( {q - 1} \right){T_{PRI}} + \left( {l - 1} \right){d_r}\sin \left( {{\theta _k}} \right)} \right)}}{{\bf{a}}^T}\left( {{\theta _k}} \right){\bf{s}}\left( \tau  \right)} \nonumber \\
  &+ {w_l}\left( {q{T_{PRI}} + \tau  + \frac{{2d}}{c}} \right),
\end{align}
where  $d$ is the distance of the range bin of interest;
   ${w_l}$ contains both interference and noise;
\begin{align}
{\bf{a}}\left( {{\theta _k}} \right) = {\left[ {1,{e^{j\frac{{2\pi }}{\lambda }{d_t}\sin \left( {{\theta _k}} \right)}},\ldots,{e^{j\frac{{2\pi }}{\lambda }\left( {{M_t} - 1} \right){d_t}\sin \left( {{\theta _k}} \right)}}} \right]^T}, \label{transmit_steering}
 \end{align}
and  ${\bf{s}}\left( \tau \right) = {\left[ {{s_1}\left( \tau \right),\ldots,{s_{{M_t}}}\left( \tau \right)} \right]^T}$.   For convenience, the signal parameters are summarized in Table \ref{table_parameter}.

\begin{table}[ht]
\caption{List of parameters used in the signal model} 
\centering 
\begin{tabular}{|l|l|}
\hline 
$d_t$ & spacing between the transmit antennas  \\ 
\hline 
$d_r$ & spacing between the receive antennas   \\ 
\hline
$M_t$ & number of transmit antennas  \\
\hline
$M_r$ & number of receive antennas  \\
\hline
$Q$ & number of pulses in a coherent processing interval  \\
\hline 
$T_{PRI}$ & radar pulse repetition interval  \\
\hline
$q$ & index of radar pulse (slow time)  \\
\hline 
$\tau$ & time in one pulse (fast time)\\
\hline
$\vartheta$ & speed of target\\
\hline
$\phi _m$ & baseband waveform\\
\hline
$d$ & distance of range bin of interest\\
\hline
$c$ & speed of light\\
\hline
$\theta$ & direction of arrival of the target\\
\hline
$\beta$ & target reflect coefficient\\
\hline
$\lambda$ & wavelength of carrier signal\\
\hline
$w_l$ & interference and white noise in the $l$th antenna\\
\hline
$T_p$ & duration of one pulse\\
\hline
$T_s$ & Nyquist sampling period\\
\hline
\end{tabular}
\label{table_parameter} 
\end{table}

\begin{figure}
\centering
\subfigure[]{\includegraphics[height=1.62in]{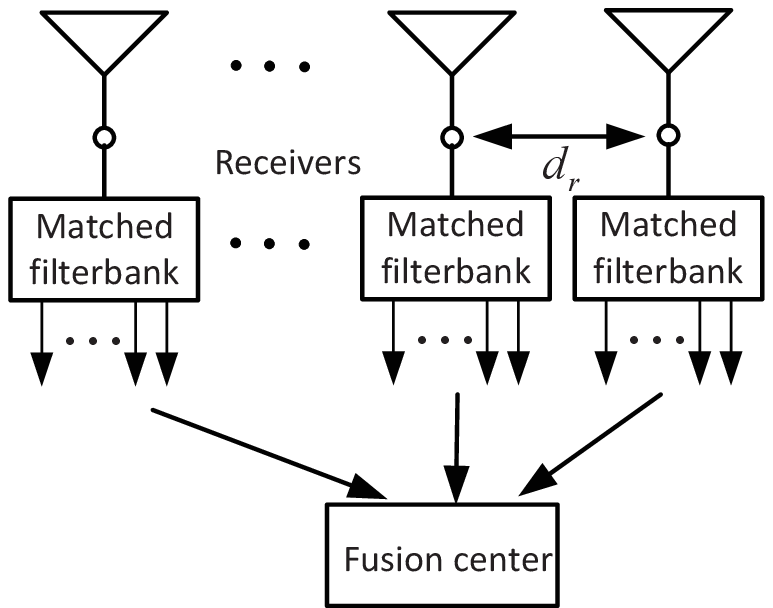}} 
\subfigure[]{\includegraphics[height=1.62in]{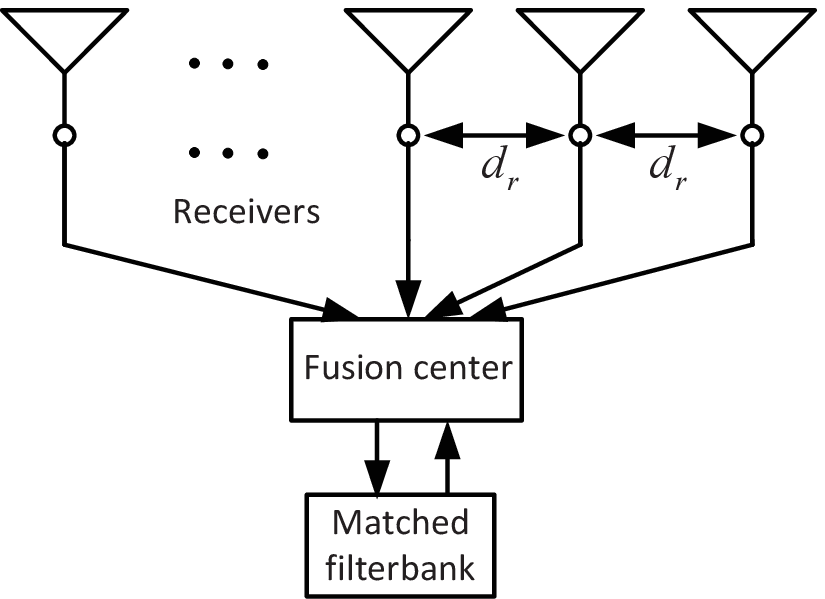}}
\caption{Two sampling schemes in the colocated MIMO radar system: (a) Sampling scheme I; (b) Sampling scheme II.} 
\label{schemes}
\end{figure}

At the $l$-th receive node, for $(l=1,\ldots,M_r)$, a matched filter bank \cite{chen} is used to extract the returns due to each transmit antenna \cite{chen} (see Fig. \ref{schemes} (a)).
Consider a filter bank  composed of $M_t$ filters, corresponding to the $M_t$  orthogonal transmit waveforms. The receive node performs  $M_t$ correlation operations and the maximum of each matched filter is forwarded to the fusion center. At the fusion center, the received signal due to the $i$-th matched filter of the $l$-th receive node, during the $q$-th pulse, can be expressed in equation (\ref{equation_scheme_I})

 \begin{align} \label{equation_scheme_I}
 {x_q(l,i)} =
\sum\limits_{k = 1}^K {{\beta _k}{e^{j\frac{{2\pi }}{\lambda }\left( {2{\vartheta _k}\left( {q - 1} \right){T_{PRI}} + \left( {l - 1} \right){d_r}\sin \left( {{\theta _k}} \right) + \left( {i - 1} \right){d_t}\sin \left( {{\theta _k}} \right)} \right)}} + w_q(l,i)}
 \end{align}
for $l=1,\ldots, M_r$, $i=1,\ldots,M_t$, and $q=1,\ldots,Q$, where ${w_q(l,i)}$ is the corresponding interference plus white noise.

 Based on the data from all receive antennas, the fusion center can construct a matrix ${\bf X}_q^{MF}$, of size
$M_r\times M_t$, whose $(l,i)$ element equals $ x_q(l,i)$.
That matrix can be expressed as
\begin{align} \label{statistics_matrix-1}
{\bf{X}}_q^{MF} = \underbrace {{\bf{B\Sigma }}{{\bf{D}}_q}{{\bf{A}}^T}}_{{\bf{Z}}_q^{MF}} + {\bf{W}}_q^{MF},
\end{align}
where ${{\bf{W}}^{MF}_q} $ is
the filtered noise; ${{\bf{D}}_q} = {\rm{diag}}\left( {{{\bf{d}}_q}} \right)$,
with ${{\bf{d}}_q} = {\left[ {{e^{j\frac{{2\pi }}{\lambda }2{\vartheta _1}\left( {q - 1} \right){T_{PRI}}}},\ldots,{e^{j\frac{{2\pi }}{\lambda }2{\vartheta _K}\left( {q - 1} \right){T_{PRI}}}}} \right]^T}$; ${\bf{\Sigma }} = {\rm{diag}}\left( {\left[ {{\beta _1},\ldots,{\beta _K}} \right]}
\right)$; ${\bf{A}}$ is the   ${M_t} \times K$  transmit steering matrix, defined as ${\bf{A}} = \left[ {{\bf{a}}\left( {{\theta _{{1}}}} \right),\ldots,{\bf{a}}\left( {{\theta _{{K}}}} \right)} \right]$;
${\bf{B}}$  is the ${M_r} \times K$ dimensional receive steering matrix, defined
in a similar fashion based on the receive steering vectors
\begin{align}
{\bf{b}}\left( {{\theta _k}} \right) = {\left[ {1,{e^{j\frac{{2\pi }}{\lambda }{d_r}\sin
\left( {{\theta _k}} \right)}},\ldots,{e^{j\frac{{2\pi }}{\lambda }\left( {{M_r} - 1}
\right){d_r}\sin \left( {{\theta _k}} \right)}}} \right]^T}. \label{receive_steering}
 \end{align}

\subsubsection{MIMO-CS Radars}
MIMO-CS radars \cite{Herman},\cite{yao} differ from conventional MIMO radars
in that they use a compressive receiver at each receive antenna to obtain a small number of samples, which are then forwarded to the fusion center (see Fig. \ref{schemes} (b)).
Let $L$ denote the number of samples that are forwarded by each receive node.
If the data forwarded   by the $l$-th antenna ($l=1,\ldots,M_r$) are inserted in the $l$-th row of an  $M_r\times L$ matrix, ${{\bf{X}}_q}$, then, an equation similar to (\ref{statistics_matrix-1}) holds, except that now the transmit
waveforms also appear in the expression, i.e.,
\cite{Sidiropoulos}
\begin{align} \label{data_matrix}
{{\bf{X}}_q} = \underbrace {{\bf{B\Sigma }}{{\bf{D}}_q}{{\bf{A}}^T}{\bf{S}}}_{{{\bf{Z}}_q}} + {{\bf{W}}_q}, 
\end{align}
where
${\bf{S}} = \left[ {{\bf{s}}\left( {0{T_s}} \right),\ldots,{\bf{s}}
\left( {\left( {L - 1} \right){T_s}} \right)} \right] \in {\mathbb C}^ {M_t \times L}$.

\section{The proposed MIMO-MC radar approach }
\label{mc_radar}

Looking at (\ref{statistics_matrix-1}),
if ${M_t} > K$ and ${M_r} > K$, both matrices ${\bf{\Sigma }}$ and ${{\bf{D}}_q}$ are rank-$K$.
Thus, the rank of the noise free  matrix ${{\bf{Z}}_q^{MF}} \in {\mathbb C}^{M_r\times M_t}$ is
$K$, which implies that matrix ${{\bf{Z}}_q^{MF}}$ is low-rank if both $M_t$ and $M_r$ are much
larger than $K$.

Similarly, looking at (\ref{data_matrix}),   both matrices ${\bf{\Sigma }}$ and ${{\bf{D}}_q}$ are rank-$K$.
The rank of matrix $\bf S$ is $\min \left\{ {{M_t},L} \right\}$. Let us assume that $L>M_t$. For ${M_t} > K$,
the rank of the noise free data matrix ${{\bf{Z}}_q} \in {\mathbb C}^{M_r\times L}$ is $K$.
 In other words, for ${M_r} \gg K$ the data matrix ${{\bf{Z}}_q}$ is low-rank.

Therefore, in both sampling schemes, assuming that the conditions (\textbf{A0}) and (\textbf{A1}) are satisfied, the fusion center matrix can be recovered from a small number of its entries. The estimated matrices corresponding to several pulses can be used to estimate the target parameters via MUSIC \cite{trees}, for example.

In the following, we leverage the low-rank property of the data matrices at the fusion center to propose a new MIMO radar approach.
Since both ${{\bf{Z}}_q}$ and ${{\bf{Z}}_q}^{MF}$ are formulated based on different sampling schemes at the receive nodes, we will study two cases, namely, sampling scheme I, which gives rise to ${{\bf{Z}}_q}^{MF}$, and sampling scheme II, which gives rise to ${{\bf{Z}}_q}$.

 \subsection{MIMO-MC with Sampling Scheme I}

\begin{figure}
\centering\includegraphics[width=5.in]{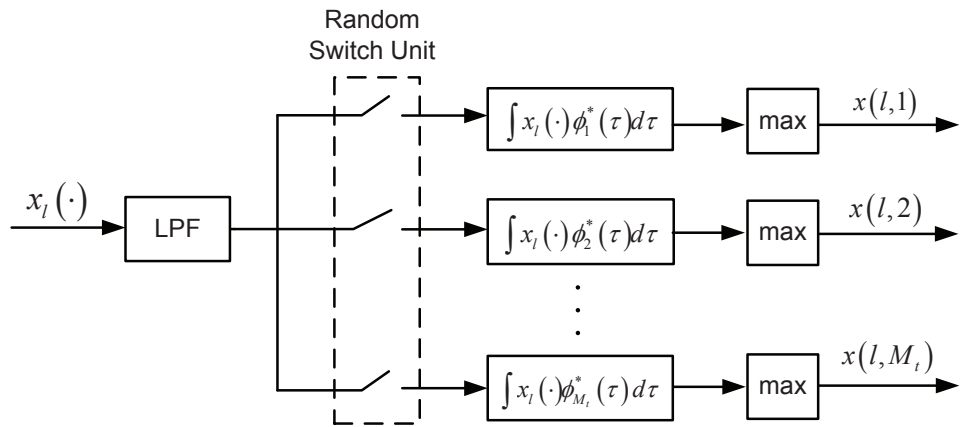}\caption{Structure of the random matched filter bank (RMFB).} \label{random_matched_filter}
\end{figure}

Suppose that the $l$th receive node uses a random matched filter bank (RMFB), as shown in Fig. \ref{random_matched_filter}, in which, a random switch unit is used to turn on and off each matched filter.
Suppose that $L_1$ matched filters are selected at random out of the $M_t$ available filters, according
to the output of a random number generator, returning  $L_1$
integers in $[0,M_t-1]$ based on the seed $s_l$. Let
 ${\mathcal J}^l$ denote the set of indices of the selected filters.
  The same random generator algorithm is also available to the fusion center. The
$l$-th receive antenna forwards the $L_1$ samples along with
the seed $s_l$ to the fusion center. Based on the seed $s_l$, the fusion center  generates the indices ${\mathcal J}^l$. Then,
it places the
 $j$-th sample of the $l$-th antenna in  the $M_r\times M_t$ matrix ${{\bf{Z}}_q}^{MF}$ at
location $(l, {\mathcal J}^l(j))$. In total, $L_1 M_t$ entries of the matrix are filled. The fusion center
declares the rest of the
entries as ``missing,'' and assuming that ${{\bf{Z}}_q}^{MF}$ meets (\textbf{A0}) and (\textbf{A1}),
applies MC techniques to estimate the full data matrix.
%

Since the samples forwarded by the receive nodes are obtained in a random sampling fashion,
the filled entries of ${\bf Z}_q^{MF}$ will correspond to a uniformly random sampling of ${\bf Z}_q^{MF}$.
In order to show that ${\bf Z}_q^{MF}$ indeed satisfies (\textbf{A0}), and as a result (\textbf{A1}),
 we need to show that the maximum   coherence of the spaces spanned by the left and
 right singular vectors of ${\bf{Z}}_q^{MF}$ is bounded by a number, $\mu_0$.
The smaller that number, the fewer samples of ${\bf{Z}}_q^{MF}$ will be required for estimating the matrix.
The theoretical analysis is pursued separately in \cite{Dennis_submitted}. Here, we confirm the applicability of MC techniques via simulations.

 We consider a scenario with  $K=2$ point targets. The DOA of the first target, $\theta_1$, is  taken to be uniformly distributed in $\left[ { - {{90}^ \circ },{{90}^ \circ }} \right]$, while the DOA of the second target is taken to be $\theta_2=\theta_1+\Delta\theta$.
The target speeds are taken to be uniformly distributed in $\left[ {0,500} \right]\rm{{m \mathord{\left/
 {\vphantom {m s}} \right.
 \kern-\nulldelimiterspace} s}}$, and the target reflectivities, $\beta_k$ are taken to be zero-mean Gaussian.
  Both the transmit and receive arrays follow the ULA model
 with $d_t=d_r=\frac{\lambda }{2}$. The  carrier frequency is taken as $f=1 \times10^9$Hz.

\begin{figure}[t]
\centering
\subfigure[]{\includegraphics[height=2.4in]{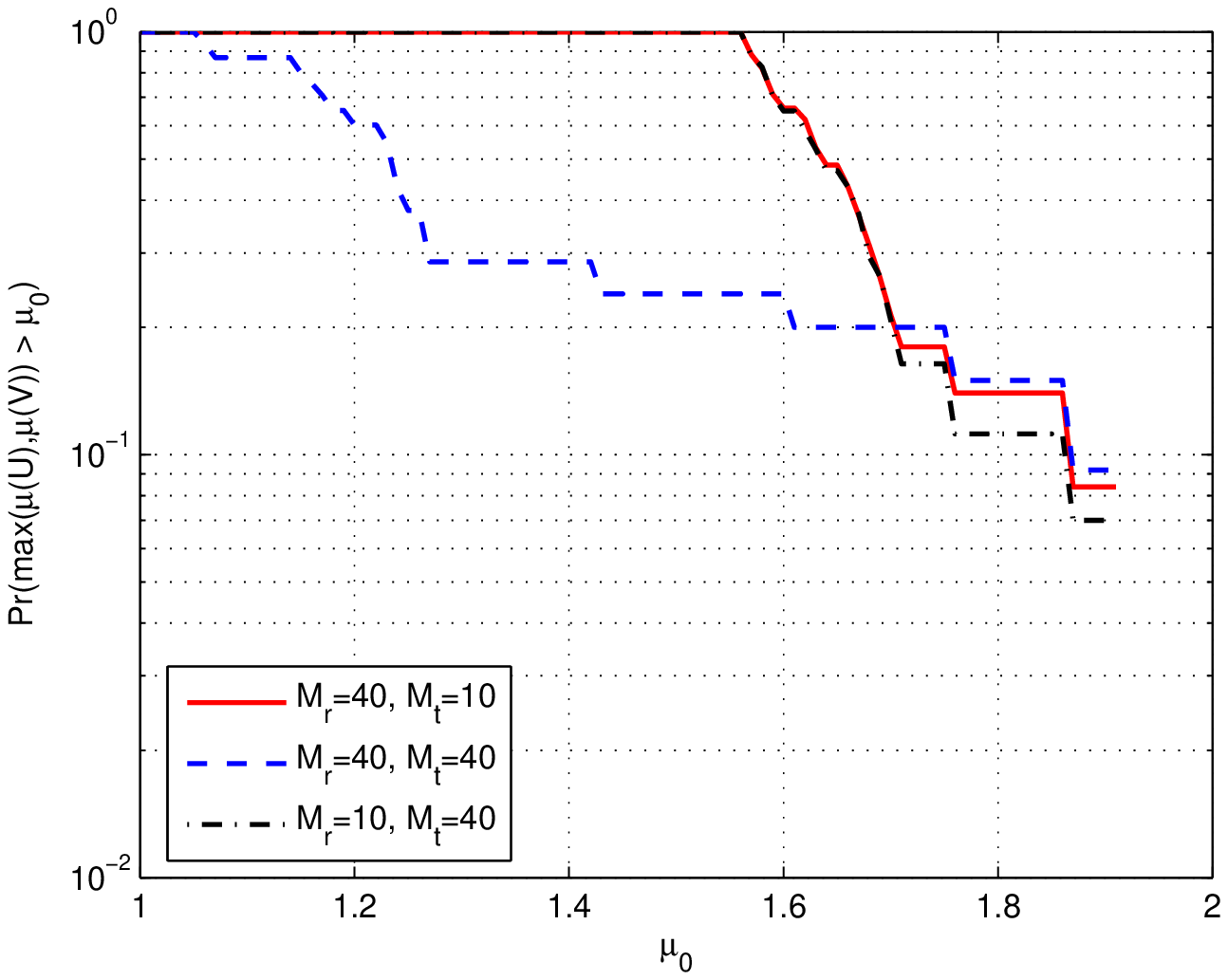}}
\subfigure[]{\includegraphics[height=2.4in]{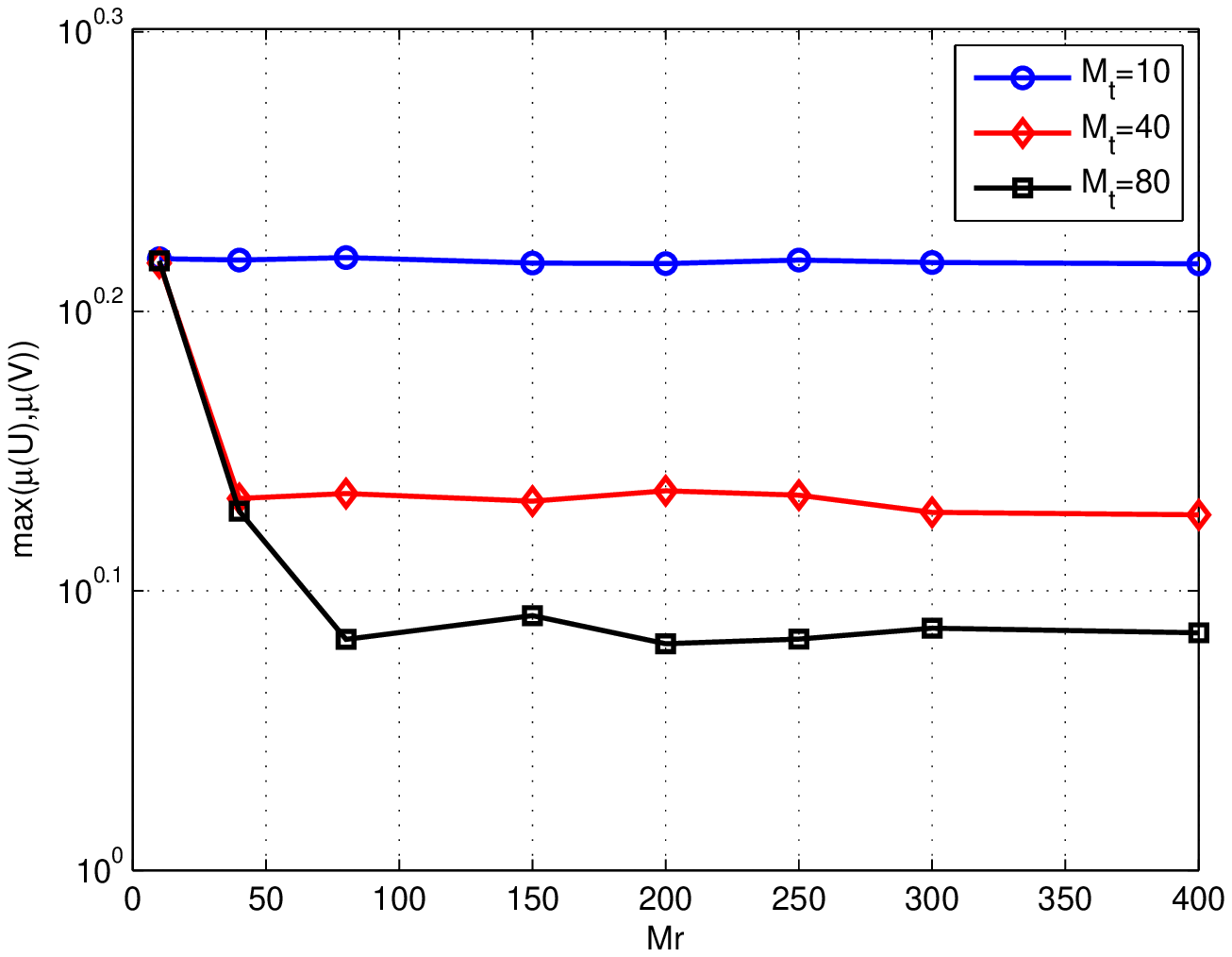}}
\subfigure[]{\includegraphics[height=2.4in]{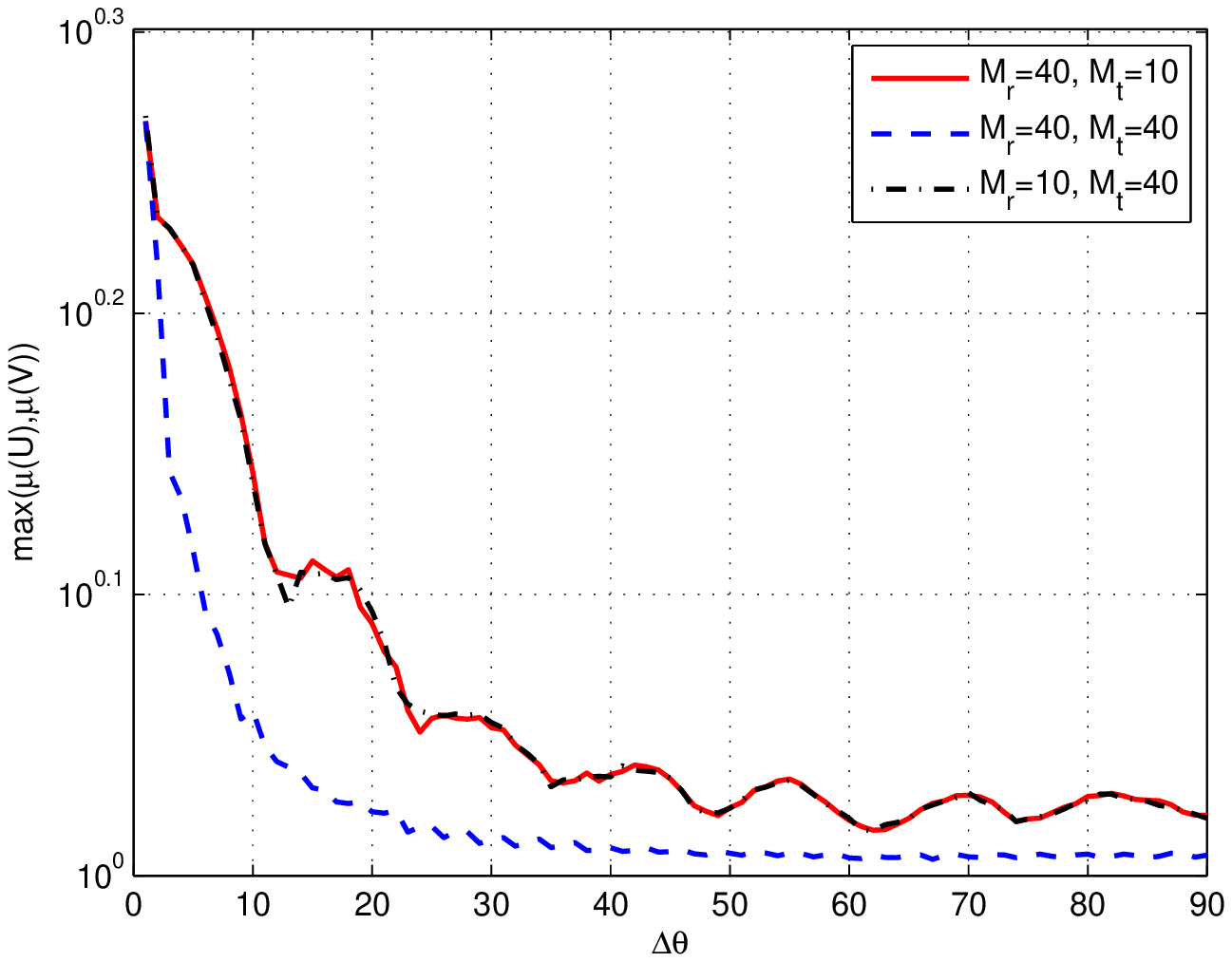}}

\caption{Scheme I, $K=2$ targets: (a) the probability of $\Pr \left( {\max \left( {\mu \left( U \right),\mu
\left( V \right)} \right) > {\mu _0}} \right)$ of ${\bf Z}_q^{MF}$ for
 $\Delta\theta=5^ \circ$;
 (b) the average ${\max \left( {\mu \left( U \right),\mu \left( V \right)} \right)}$
of ${\bf Z}_q^{MF}$  as function of number of transmit and receive antennas, and for   $\Delta\theta=5^ \circ$;
(c) the average ${\max \left( {\mu \left( U \right),\mu \left( V \right)} \right)}$
of ${\bf Z}_q^{MF}$  as function of DOA separation.   }
\label{coherence_scheme_I_prob}
\end{figure}

The left and right singular vectors of ${\mathbf Z}^{MF}_q$ were computed for $500$ independent
realizations of $\theta_1$ and target speeds. Among all the runs, the probability that
$\max \left( {\mu \left( U \right),\mu \left( V \right)} \right) > {\mu _0}$
 is shown in Fig. \ref{coherence_scheme_I_prob} (a) for $\Delta\theta=5^ \circ$
 and different values of $M_r, M_t$. One can see from the figure that in all cases, the probability
 that the coherence is bounded by a number less than $2$ is very high, while the bound gets tighter
 as the number of receive or transmit antennas increases.
On the average,  over all independent realizations, the ${\max \left( {\mu \left( U \right),\mu \left( V \right)}
  \right)}$ corresponding to  different number of receive and transmit antennas and fixed $\Delta\theta$,
   appears to
  decrease as the number of transmit and receive antennas
  increases (see Fig. \ref{coherence_scheme_I_prob} (b)). Also, the maximum
  appears to decrease  as
 $\Delta\theta$ increases, reaching $1$ for large $\Delta\theta$ (see Fig.
\ref{coherence_scheme_I_prob} (c). The rate at which the maximum reaches $1$ increases as the number of antennas increases.

%

\begin{figure}[t]
\centering
\includegraphics[height=3.in]{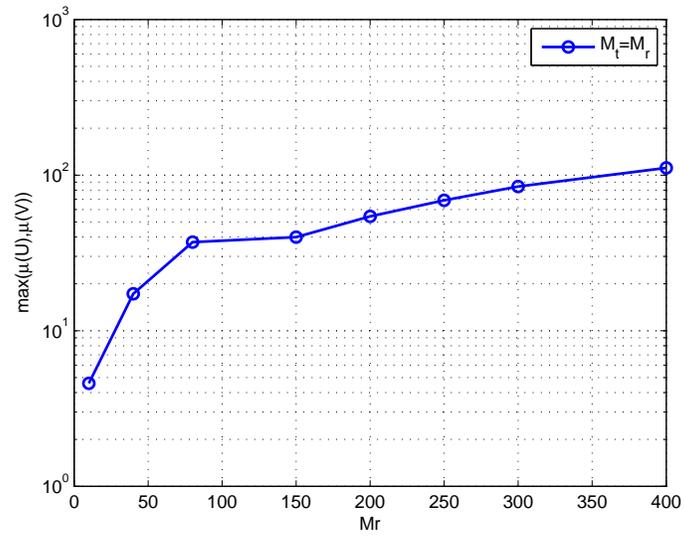}
\caption{Scheme I, $K=2$ targets: The ${\max \left( {\mu \left( U \right),\mu \left( V \right)} \right)}$
in terms of $M_r$ for  $\Delta\theta=0^\circ$, $M_t=M_r$. }  \label{coherence_scheme_I}
\end{figure}

It is interesting to see what happens at the limit $\Delta \theta=0$, i.e., when the two targets  are on
a line in the angle plane. Computing the coherence based on the assumption of rank $2$, i.e., using two eigenvectors,  the coherence shown in Fig. \ref{coherence_scheme_I} appears unbounded as $M_r$ changes. However, in this case,
the true rank of  ${\mathbf Z}_q^{MF}$  is $1$, and ${\mathbf Z}_q^{MF}$  has the best possible coherence.
Indeed, as it is shown in the  Appendix, for a rank-$1$ ${\mathbf Z}_q^{MF}$, it holds that $\mu_0=\mu_1=1$. Consequently, according to Theorem 1, the required number of entries to estimate ${\mathbf{Z}}_q^{MF}$ is minimal. This explains why in Fig. \ref{error_mc} (discussed further in Section IV) the relative recovery error of  ${\mathbf{Z}}_q^{MF}$ goes to the reciprocal of SNR faster when the two targets have the same DOA.
Of course,
   in this  case, the two targets with the same DOA appear as one, and  cannot be separated in the angle space unless other parameters, e.g., speed or range are used.
For  multiple targets, i.e., for  $K\ge 3$, if there are $n$  ($n<K$) targets with the same DOA,
 the rank of ${\bf Z}_q^{MF}$ is $K-n$, which  yields a low coherence condition since these $K-n$ DOAs
 are separated.

\subsection{MIMO-MC with Sampling Scheme II}

Suppose that the Nyquist rate samples of  signals at the receive nodes
correspond to sampling times $t_i=i T_s, \ i=0,\ldots,N-1$ with $N=T_p/T_s$.  Instead
of the receive nodes sampling at the Nyquist rate, let the
$l$-th receive antenna sample
 at times $\tau^l_j=j T_s, \ j\in {\mathcal J}^l$, where  ${\mathcal
J}^l $ is the output of a random number generator, containing $L_2$
integers in the interval $[0,N-1]$ according to a unique seed $s_l$. The
$l$-th receive antenna  forwards the $L_2$ samples along with
the seed $s_l$ to the fusion center. Under the assumption that the
fusion center and the receive nodes use the same random number generator algorithm,
the fusion center places the $j$-th sample of the $l$-th antenna in the $M_r\times N$ matrix ${\tilde {\bf{Z}}_q}$ at
location $(l, {\mathcal J}^l(j))$,  and declares the rest of the
samples as ``missing''.

The full ${\tilde {\bf{Z}}_q}$ equals:
\begin{align} \label{data_matrixII}
{{{\bf{\tilde Z}}}_q}=  {{\bf{B\Sigma }}{{\bf{D}}_q}{{\bf{A}}^T}{\tilde {\mathbf S}}} , 
\end{align}
where
$\tilde {\bf{S}} = \left[ {{\bf{s}}\left( {0{T_s}} \right),\ldots,{\bf{s}}\left( {\left( {N - 1} \right){T_s}} \right)} \right] $.
Per the discussion on ${{\bf{Z}}_q}$,  assuming that $N>M_t>K$, ${\tilde{\bf{Z}}_q}$ will be low-rank, with rank equal to $K$. Therefore, under conditions (\textbf{A0}) and (\textbf{A1}),
 ${\tilde {\bf{Z}}_q}$ can be estimated based on $m=L_2M_r$ elements, for $m$ sufficiently large.



The left singular vectors of $\tilde {\bf Z}_q$ are the eigenvectors of
${\tilde {\mathbf{Z}}_q}{\tilde {\mathbf{Z}}}_q^H  = {\mathbf{H} \tilde {\bf S}}
{{\tilde {\mathbf S}}}^H {{\bf{H}}^H}$,
where ${\bf H}={\bf{B}}{\bf{\Sigma }}{{\bf{D}}_q}{{\bf{A}}^T}$.
The right singular vectors of $\tilde {\bf Z}_q$ are the eigenvectors of
${{\tilde {\mathbf S}}^H}{{\bf{H}}^H}{\bf{H} \tilde {\bf S}}$. Since the transmit waveforms are orthogonal, it holds that
${\tilde {\mathbf S}}{{\tilde {\mathbf S}}^H} = {\mathbf{I}}$ \cite{Sidiropoulos}.
Thus, the left singular vectors are only determined by  matrix ${\bf H}$, while the right singular vectors are affected by both transmit waveforms and matrix ${\bf H}$.

Again, to check whether $\tilde {\bf Z}_q$, satisfies the conditions for MC, we resort to simulations. In particular, we show that the maximum coherence
of ${\tilde {\bf Z}}_q$ is  bounded by a small positive number ${\mu _0}$.
Assume there are $K=2$ targets. The DOA of the first target, $\theta_1$, is uniformly distributed in $\left[ { - {{90}^ \circ },{{90}^ \circ }} \right]$ and the DOA of the second target is set as $\theta_1+\Delta{\theta}$. The corresponding speeds are uniformly distributed in $\left[ {150,450} \right]{{\rm{m}} \mathord{\left/
 {\vphantom {{\rm{m}} {\rm{s}}}} \right.
 \kern-\nulldelimiterspace} {\rm{s}}}$. The target reflectivities, $\beta_k$, are zero-mean Gaussian distributed. The transmit waveforms are taken to be complex Gaussian orthogonal (G-Orth). The carrier frequency is $f= {10^9}$ Hz, resulting in $\lambda  = {c \mathord{\left/
 {\vphantom {c f}} \right. \kern-\nulldelimiterspace} f} = 0.3$ m. The inter-spacing between transmit and receive antennas is set as ${d_t} = {d_r} = {\lambda  \mathord{\left/
 {\vphantom {\lambda  2}} \right.
 \kern-\nulldelimiterspace} 2}$, respectively.

\begin{figure}[t]
\centering
\subfigure[]{\includegraphics[height=2.4in]{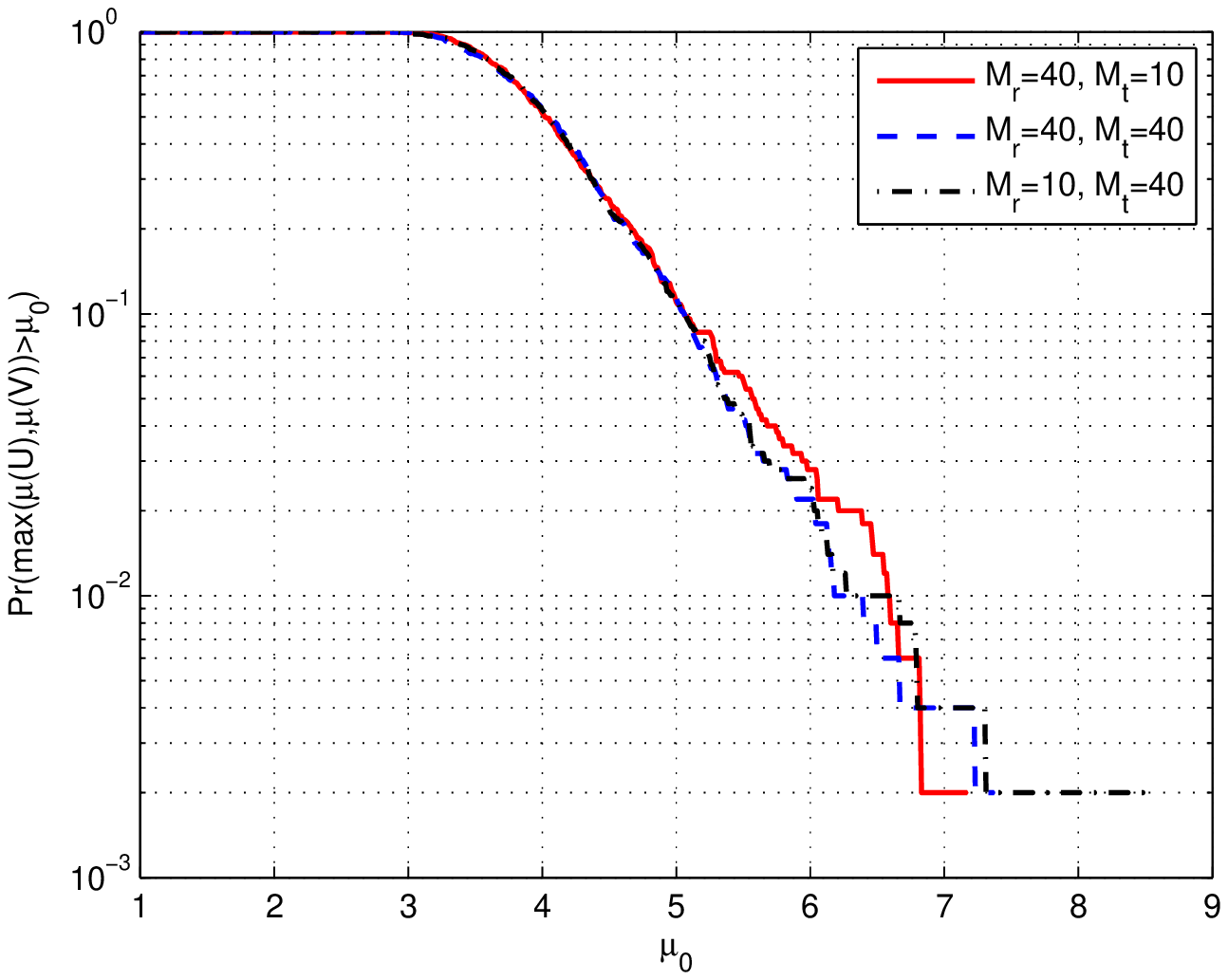}}
\subfigure[]{\includegraphics[height=2.4in]{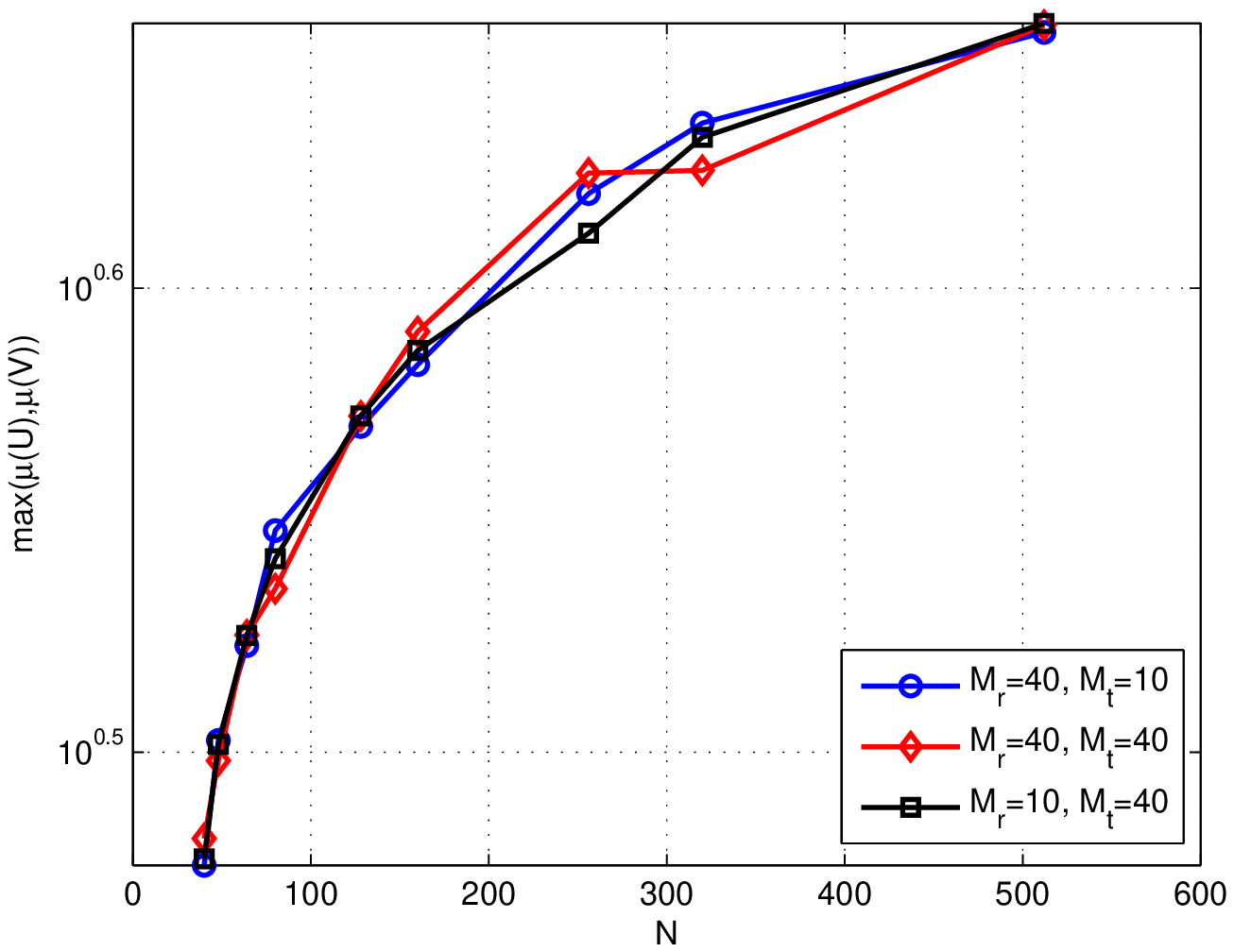}}
\subfigure[]{\includegraphics[height=2.4in]{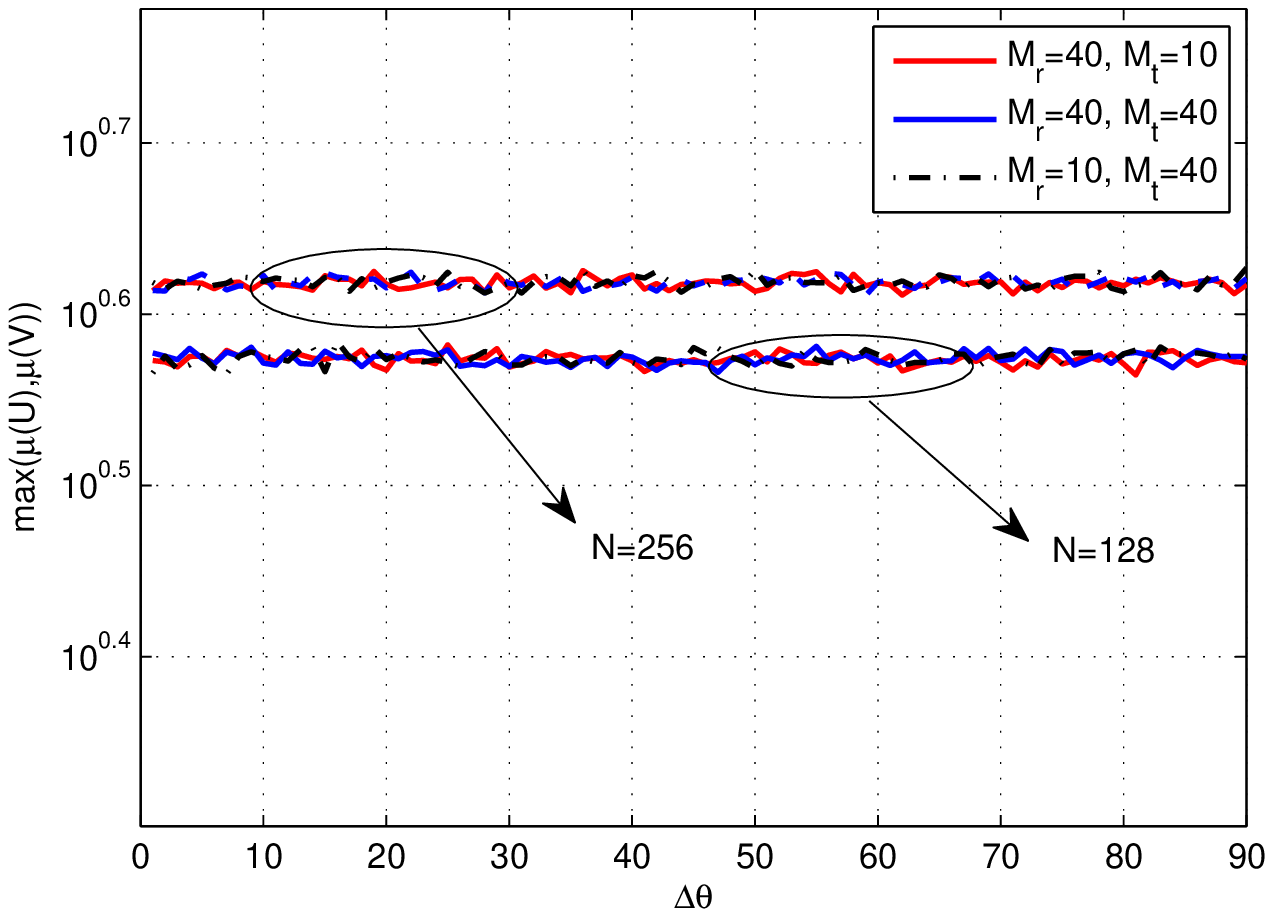}}

\caption{Scheme II, $K=2$ targets, and G-Orth  waveforms: (a) The probability of $\Pr \left( {\max \left( {\mu \left( U \right),
\mu \left( V \right)} \right) > {\mu _0}} \right)$ of $\tilde{\bf Z}_q$ for  $\Delta\theta=5^\circ$ and $N=256$;
(b) The average ${\max \left( {\mu \left( U \right),\mu \left( V \right)} \right)}$ of
 $\tilde{\bf Z}_q$ as function of $N$, for
  $\Delta\theta=5^\circ $ and different values of $M_t,M_r$;
  (c) The average ${\max \left( {\mu \left( U \right),\mu \left( V \right)} \right)}$ of
 $\tilde{\bf Z}_q$ as function of $\Delta \theta$, for
  $N=128,256$, and different combinations of $M_r,M_t$.}
    \label{coherence_scheme_II_prob}
\end{figure}

 The left and right singular vectors of $\tilde {\mathbf Z}_q$ are computed for $500$
independent realizations of $\theta_1$ and target speeds. Among all the runs, the
probability that the $\max \left( {\mu \left( U \right),\mu \left( V \right)} \right) > {\mu _0}$
is shown in Fig. \ref{coherence_scheme_II_prob}, for different values of $M_t,M_r$, $\Delta\theta=5^ \circ$, and $N=256$.
One can see from the figure that in all cases, the probability
that the coherence is bounded by a number less than $7$ is very high, while the bound gets tighter
as the number of receive or transmit antennas increases.
On  average,  over all independent realizations, the ${\max \left( {\mu \left( U \right),\mu \left( V \right)}
  \right)}$ corresponding to different values of $M_t,M_r$ and a fixed $\Delta\theta$ appears to
  increase with  $N$, (see Fig. \ref{coherence_scheme_II_prob} (b)),
  while the increase
   is not affected by the number of transmit and receive antennas.
   The average maximum does not appear to change as $\Delta\theta$ increases, and this holds for various values of
   $M_t,N$(see Fig. \ref{coherence_scheme_II_prob} (c)).

\begin{figure}
\centering
{\includegraphics[height=3.in]{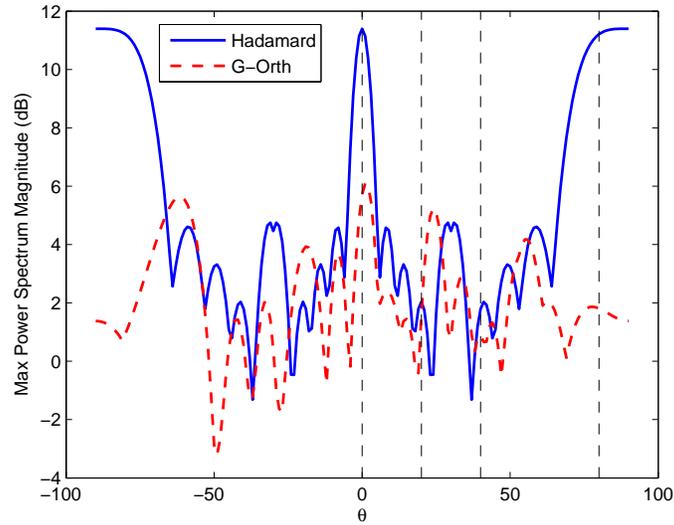}}
\caption{The maximal power spectrum of the orthogonal waveforms over $N=32$ snapshots for $M_t=10$. }
\label{power_spectrum}
\end{figure}

\begin{figure}
\centering
\subfigure[]{\includegraphics[height=2.4 in]{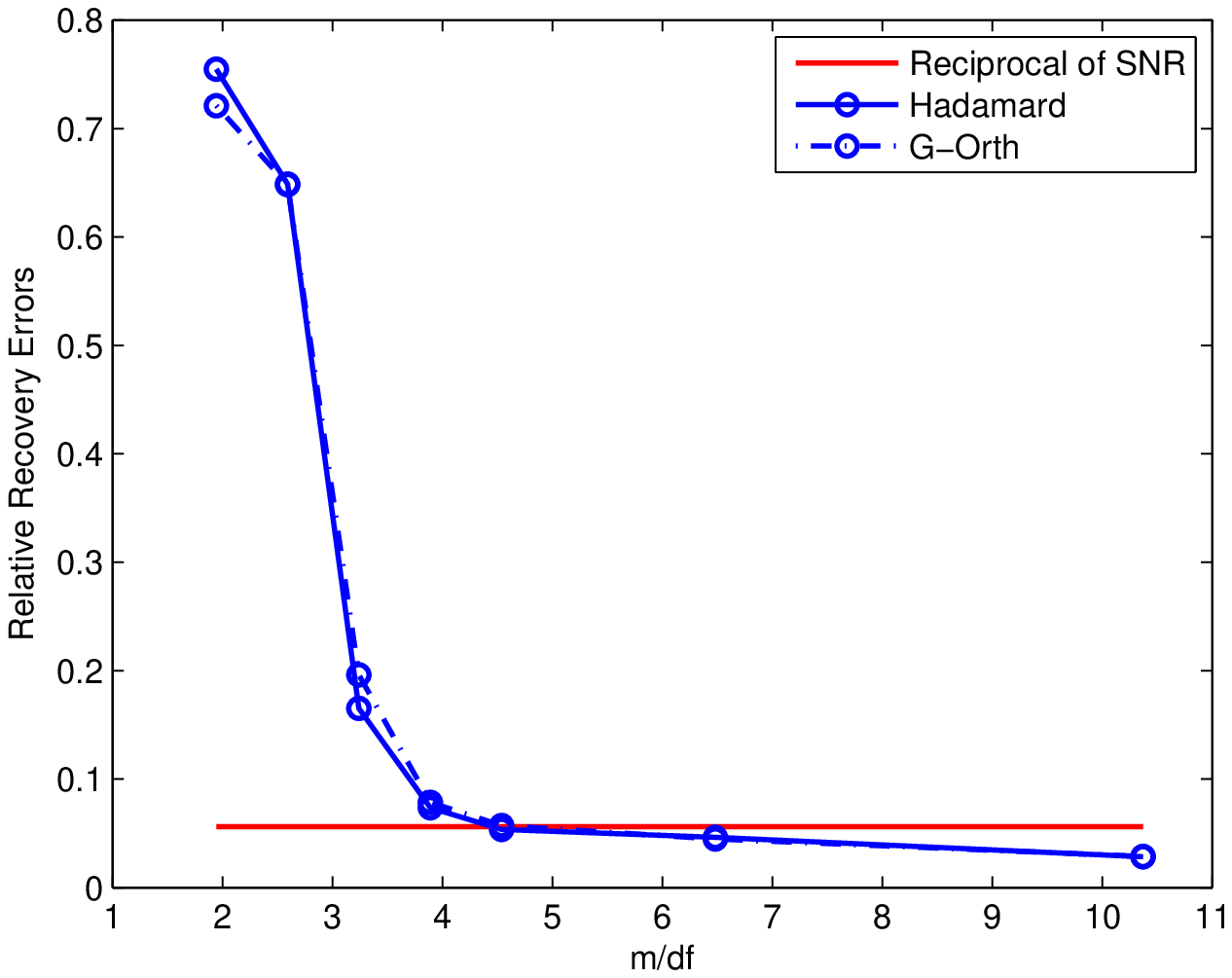}}
\subfigure[]{\includegraphics[height=2.4 in]{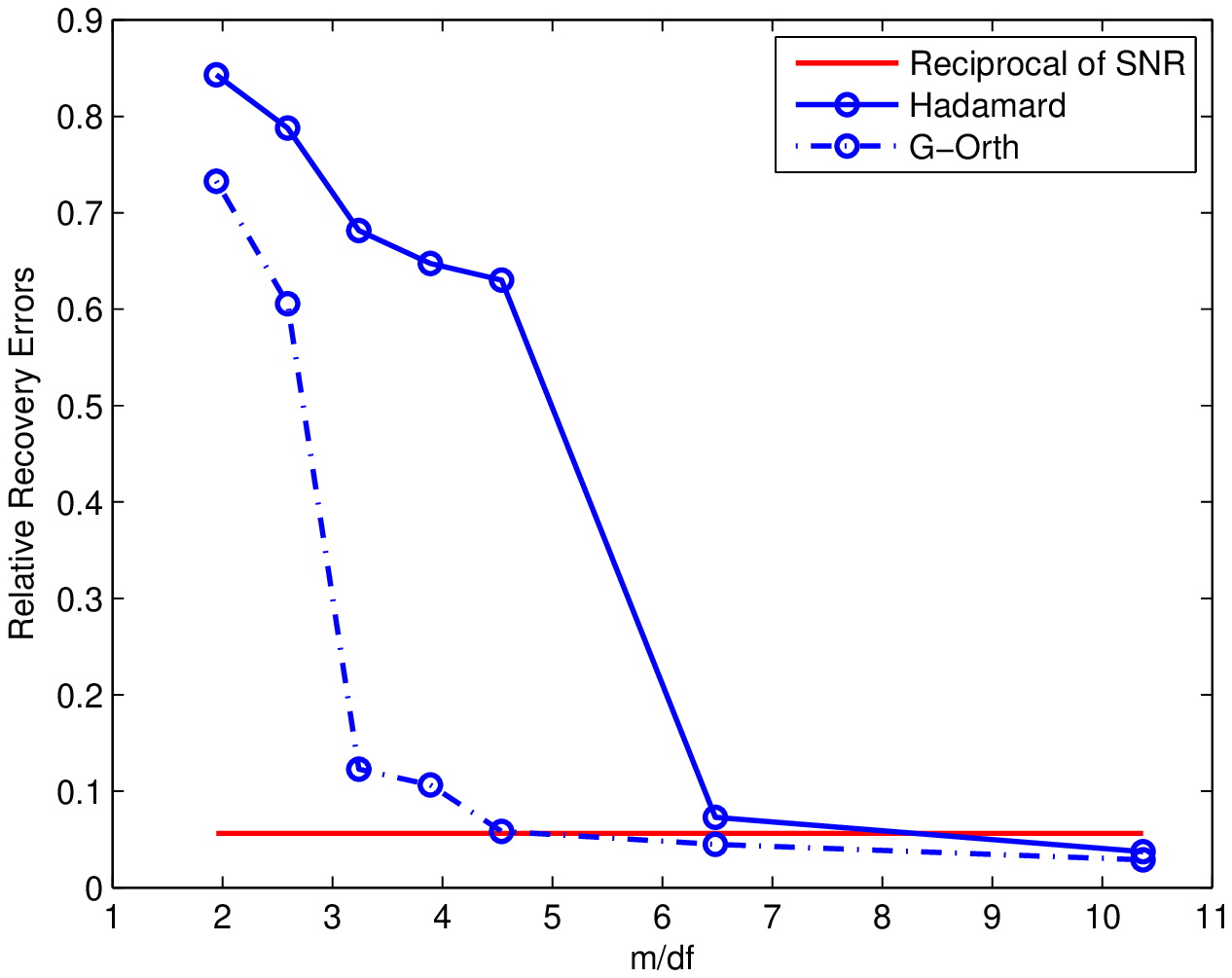}}
\caption{The comparison of matrix completion in terms of relative recovery errors with $M_r=128, M_t=10$, $N=32$, $\rm{SNR}=25\rm{dB}$.
There are $K=2$ targets located at (a) ${{20}^ \circ }$ and ${{40}^ \circ }$; (b) ${{0}^ \circ }$ and ${{80}^ \circ }$.
 }
\label{mc_error_comp_scheme_II}
\end{figure}

Based on our simulations, the MC reconstruction depends on the waveform. In particular, the coherence bound is related to the  power spectrum of each column of the waveform matrix (each column can be viewed as a waveform snapshot across the transmit antennas). Let $\tilde {S_i}\left( \omega  \right)$ denote the power spectrum of the $i$-th column of $\tilde {\bf S }\in {{\mathbb{C}}^{{M_t} \times N}}$. If $\tilde {S_i}\left( \omega  \right)$ is similar for different $i$'s,
 the MC recovery performance improves with increasing $M_t$ (or equivalently, the coherence bound decreases) and does not depend on $N$; otherwise, the performance worsens with increasing  $N$ (i.e., the coherence bound increases). When the $\tilde {S_i}\left( \omega  \right)$  has peaks at certain $\omega$'s  that occur close to  targets, the performance worsens.
In Fig. \ref{power_spectrum}, we show the maximum power spectra values corresponding to Hadamard and G-Orth waveforms  for $M_t=10$ and $N=32$. It can be seen in Fig. \ref{power_spectrum} that the maximum power spectrum values corresponding to the Hadamard waveform have strong peaks at certain $\omega$'s, while those for  the G-Orth waveforms fluctuate around a low value.
Suppose that there are two targets  at angles  $\theta_1={{20}^ \circ }$ and ${\theta_2={40}^ \circ }$,  corresponding to ${\omega _1} = \frac{1}{2}\sin \left( {\frac{\pi }{9}} \right)$ and $\omega_2=\frac{1}{2}\sin \left( {\frac{{2\pi }}{9}} \right)$, respectively.
From Fig. \ref{power_spectrum} one can see that the  targets fall under low power spectral values for both waveform cases. The corresponding MC recovery error, computed based on $50$ independent runs is shown in Fig. \ref{mc_error_comp_scheme_II} (a). One can see that the error  is the same for both  waveforms.
As another case, suppose that the two targets are at angles
 ${{0}^ \circ },{{80}^ \circ }$, corresponding to  ${\omega _1} = 0, \omega_2=\frac{1}{2}\sin \left( {\frac{{4\pi }}{9}} \right)$, respectively.
  Based on Fig. \ref{power_spectrum}, one can see that $\omega_1$ and $\omega_2$ fall under high spectral peaks in the case of Hadamard waveforms. The corresponding MC recovery error is shown in Fig. \ref{mc_error_comp_scheme_II}(b), where one can see that Hadamard waveforms yield higher error.

\subsection{Discussion of MC in Sampling Schemes I and II}
To apply the matrix completion techniques in colocated MIMO radar, the data matrices $ \tilde {\bf Z}_q \in {\mathbb C}^{M_r \times N}$ and  ${\bf Z}_q^{MF}\in {\mathbb C}^{M_r \times M_t}$ need to be low-rank, and   satisfy the coherence conditions with small $\mu_i, i=0,1$. 

We have already shown that the rank of the above two matrices equals the number of targets. In sampling scheme I, to ensure that matrix ${\bf Z}_q^{MF}$ is low-rank, both $M_t$ and $M_r$ need to be much larger than $K$, in other words, a large transmit as well as a large receive array are required.
This, along with the fact that each receiver needs a filter bank, make scheme I more expensive in terms of hardware. However,  the matched filtering operation improves the SNR in the received signals. Although in this paper we use the ULA model to illustrate the idea of MIMO-MC radar, the idea can be extended to arbitrary antenna configurations. One possible scenario with a large number of antennas is a networked radar system \cite{Dutta}\cite{Liang}, in which the antennas are placed on the nodes of a network. In such scenarios, a large number of collocated or widely separated sensors could be deployed to collaboratively perform target detection.

In sampling scheme II, assuming that more samples ($N$) are obtained than existing targets ($K$),
$ \tilde{\bf Z}_q$ will be  low-rank as long as there are more receive antennas than targets, i.e.,  $M_r\gg K$.
 For this scheme, there is no condition on the number of transmit antennas $M_t$ if G-Orth waveform is applied.

 Based on  Figs.  \ref{coherence_scheme_I_prob} and  \ref{coherence_scheme_II_prob}, it appears that
 the average coherence bound, $\mu_0$, corresponding to  $\tilde{\bf Z}_q$ is larger than that of ${\bf Z}_q^{MF}$.
 This indicates that the coherence under scheme II is larger than that under scheme I,
 which means that for scheme II,  more observations  at the fusion center are required  to recover the data matrix with missing entries.


\subsection{Target Parameters Estimation with Subspace Methods}
In this section we describe the MUSIC-based method that will be applied to the
estimated data matrices at the fusion center to yield target information.

Let  ${\bf \hat Z}_q$ denote the estimated data matrix for sampling
scheme II, during pulse $q$.
Let us perform matched filtering  on  ${\bf \hat Z}_q$ to obtain
\begin{align}
{{\bf{Y}}_q} = \frac{1}{L}{{\bf{ \hat Z}}_q}{\tilde {\bf{S}}^H} = {\bf{B}}{\bf{\Sigma }}{{\bf{D}}_q}{{\bf{A}}^T} + {{{\bf{\tilde W}}}_q},
\end{align}
where ${{{\bf{\tilde W}}}_q}$ is noise  whose distribution is a function of
the additive noise and the nuclear norm minimization problem in
(\ref{robust_mc}). For sampling scheme I, a similar equation holds
for the recovered matrix without further matched filtering.

Then, let us stack the  matrices  into vector  ${{\mathbf{y}}_q} = vec\left( {{{\mathbf{Y}}_q}} \right)$,
for sampling scheme II, or ${{\mathbf{y}}_q} = vec\left( {{\mathbf{\hat Z}}_q^{MF}} \right)$, for
 sampling scheme I. Based on
 $Q$ pulses, the following matrix can be formed:
  ${{\bf{Y}}} = \left[ {{{\bf{y}}_1},\ldots,{{\bf{y}}_Q}} \right]  \in {{\mathbb C}^{{M_t}{M_r} \times Q}}$, for which it holds that
\begin{align} \label{joint_Y}
{\bf{Y}} = {\bf{V}}\left( \theta  \right){\bf{\tilde X}} + {\bf{W}},
\end{align}
where ${\bf{\tilde X}} = \left[ {{{{\bf{\tilde x}}}_1},\ldots,{{{\bf{\tilde x}}}_Q}} \right]$
 is a $K\times Q$ matrix containing target reflect coefficient and Doppler shift information;
 ${{{\bf{\tilde x}}}_q} = {\left[ {{{\tilde x}_{1,q}},\ldots,{{\tilde x}_{K,q}}} \right]^T}$
 and ${{\tilde x}_{k,q}} = {\beta _k}{e^{j\frac{{2\pi }}{\lambda }2{\vartheta _k}
 \left( {q - 1} \right){T_{PRI}}}}$;  ${\bf{V}}\left( \theta  \right) = \left[ {{\bf{v}}\left( {{\theta _1}} \right),\ldots,{\bf{v}}\left( {{\theta _K}} \right)} \right]$
  is a $M_t M_r\times K$ matrix with columns
\begin{align}
{\bf{v}}\left( \theta  \right) = {\bf{a}}\left( \theta  \right) \otimes {\bf{b}}\left( \theta  \right)
\end{align}
and  ${\bf{W}} = \left[ {vec\left( {{{{\bf{\tilde W}}}_1}} \right),\ldots,vec\left( {{{{\bf{\tilde W}}}_Q}} \right)} \right]$.

The sample covariance matrix can be obtained as
\begin{align}
{\bf{\hat R}} = \frac{1}{Q}\sum\limits_{n = 1}^Q {{{\bf{y}}_n}{\bf{y}}_n^H = } \frac{1}{Q}{\bf{Y}}{{\bf{Y}}^H}.
\end{align}
According to  \cite{trees}, the pseudo-spectrum of MUSIC estimator can be written as
\begin{align}
P\left( \theta  \right) = \frac{1}{{{{\bf{v}}^H}\left( \theta  \right){{\bf{E}}_n}{\bf{E}}_n^H{\bf{v}}\left( \theta  \right)}} \label{spectrum}
\end{align}
where ${{\bf{E}}_n}$ is a matrix containing the eigenvectors of the noise subspace of ${\bf{\hat R}} $.
The DOAs of target can be obtained by finding the peak locations of the pseudo-spectrum (\ref{spectrum}).

For joint DOA and speed estimation, we  reshape ${\bf Y}$ into ${\bf{\tilde Y}} \in {{\mathbb{C}}^{Q{M_t} \times {M_r}}}$ and get
\begin{align}
{\bf{\tilde Y}} ={\bf{F}}{\bf{\Sigma }}\left[ {{\bf{b}}\left( {{\theta _1}} \right),\ldots,{\bf{b}}\left( {{\theta _K}} \right)} \right] + {\bf{W}},
\end{align}
where ${\bf{F}}=\left[ {{\bf{d}}\left( {{\vartheta_1}} \right) \otimes
{\bf{a}}\left( {{\theta _1}} \right),\ldots,{\bf{d}}\left( {{\vartheta_K}}
\right) \otimes {\bf{a}}\left( {{\theta _K}} \right)} \right]$,
${\bf d}\left( \vartheta  \right) = {\left[ {1,{e^{j\frac{{2\pi }}{\lambda }2\vartheta {T_{PRI}}}},\ldots,{e^{j\frac{{2\pi }}{\lambda }2\vartheta \left( {Q - 1} \right){T_{PRI}}}}} \right]^T}$. The sampled covariance matrix of the receive data signal can then be
obtained as ${{\bf{ \hat R}}_{\tilde Y}} = \frac{1}{M_r}{\bf{\tilde Y}}{{\bf{\tilde Y}}^H}$, based
on which DOA and speed joint estimation can be implemented using 2D-MUSIC. The pseudo-spectrum of 2D-MUSIC estimator is
\begin{align} \label{2D-MUSIC}
P\left( {\theta ,\vartheta } \right) = \frac{1}{{{{\left[ {{\bf{d}}\left( \vartheta  \right) \otimes {\bf{a}}\left( \theta  \right)} \right]}^H}{{\bf{E}}_n}{\bf{E}}_n^H\left[ {{\bf{d}}\left( \vartheta  \right) \otimes {\bf{a}}\left( \theta  \right)} \right]}}
\end{align}
where ${{\bf{E}}_n} \in {{\mathbb C}^{Q{M_t} \times \left( {Q{M_t} - K} \right)}}$ is the matrix constructed by the eigenvectors corresponding to the noise-subspace of ${{\bf{ \hat R}}_{\tilde Y}} $.

\section{Numerical Results}
\label{numerical}
In this section we demonstrate the performance of the proposed approaches in
terms of  matrix recovery error and DOA resolution.

We use ULAs for both transmitters and receivers.
The inter-node distance for the transmit array  is set to ${M_r}{\lambda  \mathord{\left/
 {\vphantom {\lambda  2}} \right.
 \kern-\nulldelimiterspace} 2}$, while for the  receive antennas is set as ${\lambda  \mathord{\left/
 {\vphantom {\lambda  2}} \right.
 \kern-\nulldelimiterspace} 2}$.
Therefore, the degrees of freedom of the MIMO radars is $M_r M_t$ \cite{Jian}, i.e., a high resolution could be achieved with a small number of transmit and receive antennas.
 The carrier frequency is set to $f = 1 \times {10^9}\rm{Hz}$, which is a typical radar frequency.
  The noise introduced in both sampling schemes
 is  white Gaussian with zero mean and variance  $\sigma^2$.
 The data matrix recovery is done using the singular value thresholding (SVT) algorithm \cite{SVT}. Nuclear norm optimization is a convex optimization problem. There are several algorithms available to solve this problem, such as TFOCS \cite{TFOCS}. Here, we chose the SVT algorithm because it is a simple first order method and is suitable for a large size problem with a low-rank solution. During every iteration of SVT, the storage space is minimal and computation cost is low.

We should note that in the SVT algorithm, the matrix rank, or equivalently, the number of targets,  is not required to be known a prior.  The only requirement is that the number of targets is much smaller than the number of TX/RX antennas, so  that the receive data matrix is low-rank. To make sure the iteration sequences of SVT algorithm converge to the solution of the nuclear norm optimization problem, the thresholding parameter $\tau$ should be large enough. In the simulation, $\tau$ is chosen empirically and set to $\tau=5\zeta $, where $\zeta$ is the dimension of the low-rank matrix that needs to be recovered.
\subsection{Matrix Recovery Error under Noisy Observations} \label{mc_revovery}

We consider a scenario with two targets. The first target DOA, $\theta_1$ is
generated at random in $\left[ { - {{90}^ \circ },{{90}^ \circ }} \right]$, and the second target DOA, is taken as $\theta_2=\theta_1+\Delta \theta$.
The target reflection coefficients are set as complex random, and the corresponding speeds are taken at random in $\left[ {0,500} \right]{{\rm{m}} \mathord{\left/
 {\vphantom {{\rm{m}} {\rm{s}}}} \right.
 \kern-\nulldelimiterspace} {\rm{s}}}$.
 The SNR at each receive antenna is set to $25\rm{dB}$.

In the following, we compute the matrix recovery error as function of the number
 of samples, $m$, per degrees of freedom, $\rm{df}$, i.e., ${m \mathord{\left/
 {\vphantom {m {{\rm{df}}}}} \right.
 \kern-\nulldelimiterspace} {{\rm{df}}}}$, a quantity
 also used in \cite{completion2}. A matrix of size $n_1\times n_2$ with rank $r$,
 has $r\left( {{n_1} + {n_2} - r} \right)$ degrees of freedom \cite{Candes&Recht2009}.
 Let ${\phi _{\mathbf{ {\hat Z}}}}$ denote the relative matrix recovery error, defined as:
 \begin{equation}
 {\phi _{\bf{{\hat Z}}}}= {{{{\left\| {{\bf{\hat Z}} - {\bf{Z}}} \right\|}_F}} \mathord{\left/
 {\vphantom {{{{\left\| {{\bf{\hat Z}} - {\bf{Z}}} \right\|}_F}} {{{\left\| {\bf{Z}} \right\|}_F}}}} \right.
 \kern-\nulldelimiterspace} {{{\left\| {\bf{Z}} \right\|}_F}}},
 \end{equation}
 where
 we use $\bf Z$  to denote the data matrix in both sampling schemes,
  and  $\bf{\hat Z}$  to denote the estimated data matrix.

\begin{figure}
\centering\includegraphics[height=3in]{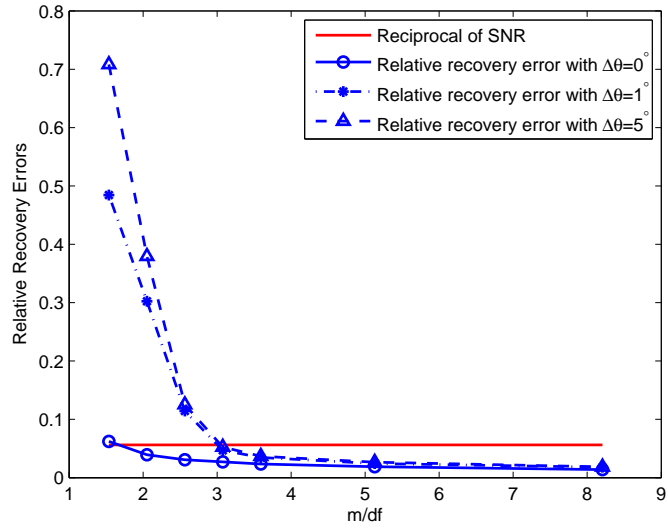}\caption{Scheme I, $K=2$ targets: the relative recovery error for ${\bf Z}_q^{MF}$   under different values of DOA separation. $M_r=M_t=40$. } \label{error_mc}
\end{figure}

 Figure \ref{error_mc} shows  ${\phi _{\mathbf{ {\hat Z}}}}$
 under sampling scheme I, versus the number of samples per
 degree of freedom for the same scenario as above. The number of transmit/receiver antennas is set as $M_t=M_r=40$.
 It can be seen from Fig. \ref{error_mc} that when ${m \mathord{\left/
 {\vphantom {m {{\rm{df}}}}} \right.
 \kern-\nulldelimiterspace} {{\rm{df}}}}$ increases from $2$ to $4$, or correspondingly,
 the matrix occupancy ratio increases from $p_1\approx0.2$ to $\approx0.4$, the relative error ${\phi _{\mathbf{{\hat Z}}}}$ drops sharply to the reciprocal of
 the  matched filter SNR level, i.e., a ``phase transition'' \cite{Keshavan} occurs.
It can be seen in Fig. \ref{error_mc} that, when the two targets have the same DOA, the relative recovery error is the smallest. This is because in that case the data matrix has the optimum coherence parameter,
i.e.,  $\mu_0=1$. As the DOA separation  between the two target increases, the relative recovery error of the data matrix in the transition phase increases.
In the subsequent DOA resolution  simulations, we set the matrix occupancy ratio as
$p_1 =\frac{L_1 M_r}{M_t M_r}=0.5$, which corresponds to  ${m \mathord{\left/
 {\vphantom {m {{\rm{df}}}}} \right.
 \kern-\nulldelimiterspace} {{\rm{df}}}} \approx 5$, to ensure that the relative recovery error
 has dropped to the reciprocal of SNR level.

\begin{figure}
\centering\includegraphics[height=3in]{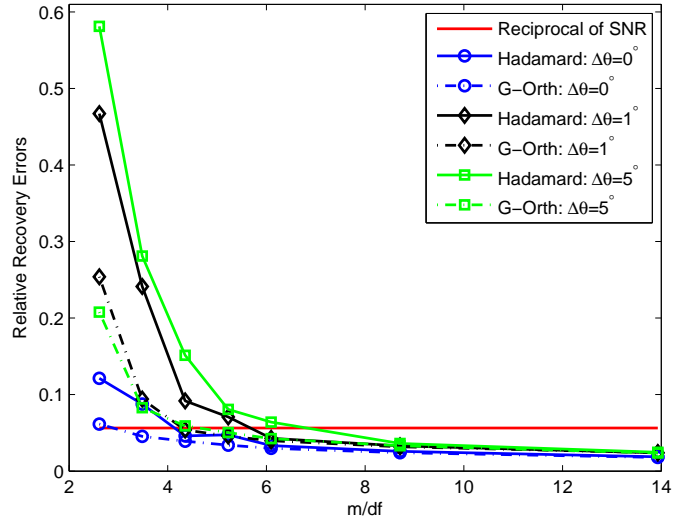}\caption{Scheme II,  $K=2$ targets: the relative recovery errors for $ \tilde{\bf Z}_q$  under Hadamard and Gaussian Orthogonal waveforms, and different values of $\Delta \theta$.  $M_r=M_t=40$, $N=256$.} \label{error_without_MF}
\end{figure}

 Figure \ref{error_without_MF} shows the relative recovery errors, $ {{\phi _{\mathbf{{\hat Z}}}}}$, for data
matrix $\tilde{\bf  Z}_q$ (sampling scheme II), corresponding to  Hadamard or Gaussian orthogonal
(G-Orth) transmit waveforms, and the number of Nyquist samples is taken to be $N=256$.
Different values of DOA separation for the two targets are considered, i.e.,  $\Delta\theta=0^{\circ},1^{\circ},5^{\circ}$, respectively.

 The results are averaged over $100$ independent angle and speed realizations; in each realization
 the $L_2$ samples are obtained at random among the $N$ Nyquist samples at each receive antenna.
The results of Fig. \ref{error_without_MF}  indicate that, for the same $\Delta\theta$, as ${m \mathord{\left/
 {\vphantom {m {{\rm{df}}}}} \right.
 \kern-\nulldelimiterspace} {{\rm{df}}}}$ increases,  the relative recovery error, $ {{\phi _{\mathbf{{\hat Z}}}}}$, under Gaussian orthogonal waveforms (dash lines) reduces to the reciprocal of the
SNR faster than  under Hadamard waveforms (solid lines). A plausible reason for this is
that under  G-Orth waveforms, the average coherence parameter of $\tilde{\bf Z}_q$
is smaller as compared with that under  Hadamard waveforms.
 Under Gaussian orthogonal waveforms,
the error $ {{\phi _{\mathbf{{\hat Z}}}}}$ decreases as  $\Delta\theta$ increases. On the other hand, for Hadamard
waveforms the relative recovery error appears to increase with an increasing $\Delta\theta$, a behavior that diminishes in the region to the right of the point of ``phase transition''.
However, the behavior of the error at the left of the ``phase transition'' point is not of interest as the matrix completion errors are  pretty high and DOA estimation is simply not possible. At the right of the ``phase transition'' point, the  observation noise dominates in the DOA estimation performance.

In both waveforms, the minimum error is achieved when $\Delta\theta=0^\circ$, i.e., when the two targets have the same DOA, in which case the rank of data matrix $\tilde{\bf Z}_q$ is rank-$1$. The above observations suggest that  the waveforms do affect performance, and optimal waveform design would be an interesting problem. The waveform selection problem could be formulated as an optimization problem under the orthogonal and narrow-band constraints. We plan to pursue this in our future work.

It can be seen from Fig. \ref{error_mc} and Fig. \ref{error_without_MF}
that in the noisy cases, as the matrix occupancy ratio increases,
the relative recovery errors of the matrices decreases to the reciprocal of SNR.

\subsection{DOA Resolution with Matrix Completion}

In this section we study the probability that two DOAs will be resolved based on the proposed techniques.
Two targets are generated at $10^\circ$ and $10^\circ+\Delta\theta$, where  ${\Delta\theta } = \left[ {{{0.05}^ \circ },{{0.08}^ \circ
},{{0.1}^ \circ },{{0.12}^ \circ },{{0.15}^ \circ },{0.18^ \circ },{0.2^ \circ },{0.22^ \circ },{0.25^ \circ },{0.3^ \circ }} \right]$. The
corresponding target speeds are set to $150$ and $400$ ${{\rm{m}} \mathord{\left/
 {\vphantom {{\rm{m}} {\rm{s}}}} \right.
 \kern-\nulldelimiterspace} {\rm{s}}}$. We set $M_t=M_r=20$ and $Q=5$.
  The DOA information is obtained by finding the peak locations of the pseudo-spectrum (\ref{spectrum}).
  If the DOA estimates ${\hat \theta }_i$, $i=1,2$ satisfy $\left| {\theta_i  - {\hat \theta}_i } \right| \le \varepsilon {\Delta\theta }, \varepsilon  = 0.1$, we declare the estimation a success.  The probability of DOA resolution is then defined as the fraction of successful events in $200$ iterations. For comparison, we also plot the probability curves with full data matrix observations.

\begin{figure}
\centering
{\includegraphics[height=3.in]{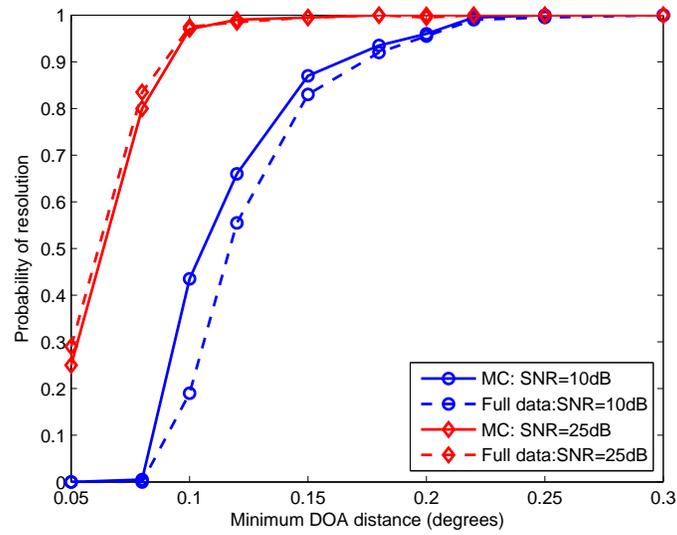}}
\caption{Scheme I:  DOA resolution. The parameter are set as $M_r=M_t=20$, $p_1=0.5$ and
$\rm{SNR}=10,25\rm{dB}$, respectively.}
\label{DOA_resolution_I}
\end{figure}

 First, for scheme I, $L_1=10$ matched filers are independently
 selected at random at each receive antenna, resulting matrix occupancy ratio of $p_1=0.5$.
 The corresponding probability of DOA resolution is shown in Fig. \ref{DOA_resolution_I} (a).
 As expected, the probability of DOA resolution
  increase as the SNR increases.
 The performance of DOA resolution based on the full set of observations has similar behavior.
 When $\rm{SNR}=25\rm{dB}$, the performance of MC-based DOA estimation is close to that with the
 full data matrix.
 Interestingly, for $\rm{SNR}=10\rm{dB}$, the MC-based result has better performance than that
 corresponding to a full data matrix. Most likely, the MC acts like a low-rank approximation of
 ${\bf { Z}}_q^{MF}$, and
 thus eliminates some of the noise.

\begin{figure}
\centering
\subfigure[]{\includegraphics[height=3.in]{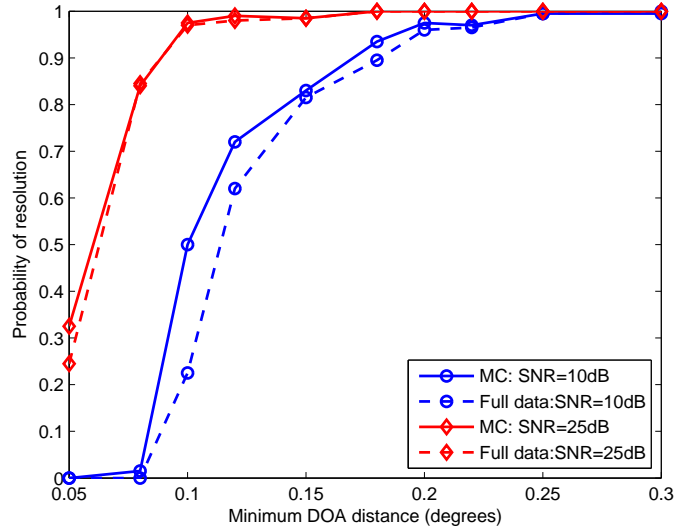}}
\subfigure[]{\includegraphics[height=3.in]{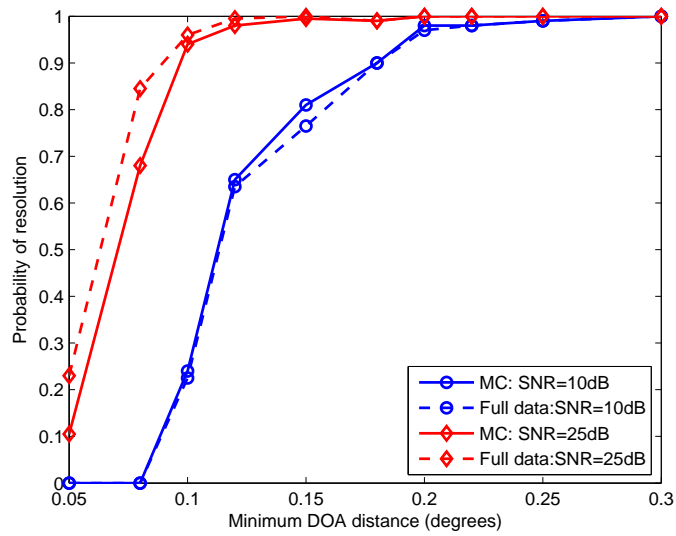}}
\caption{Scheme II, $K=2$,  $M_r=M_t=20$, $N=256$, $p_2=0.5$, $\rm{SNR}=10,25\rm{dB}$.
DOA resolution with (a) G-Orth waveforms; (b) with Hadamard waveforms.
 }
\label{DOA_resolution_II}
\end{figure}

 The probabilities of DOA resolution of DOA estimates under scheme II,
  with  G-Orth and Hadamard waveforms are plotted in Fig.
  \ref{DOA_resolution_II} (a) and (b), respectively.
  The parameters are set as $N=256$ and $p_2=0.5$, i.e.,
  each receive antenna uniformly selects $L_2=128$ samples at random to forward.
  Similarly, the simulation results show that under scheme II, the performance
  at $\rm{SNR}=10\rm{dB}$ is slightly better than that with full data access.
  In addition, it can be seen that the performance with
  G-Orth waveforms is better than  with Hadamard waveforms.
  This is because the average coherence of ${\bf Z}_q$ under Hadamard waveforms
  is higher than that with G-Orth waveforms. As shown in Fig. \ref{DOA_resolution_II}, increasing the SNR from $10\rm{dB}$ to $25\rm{dB}$ can greatly improve the DOA estimation performance, as it benefits both the matrix completion and the  performance of subspace based DOA estimation method, i.e., MUSIC (see chapt. 9 in [29]).

\subsection{Comparisons of Sampling Schemes I and II}

\begin{figure}
\centering\includegraphics[width=6.in]{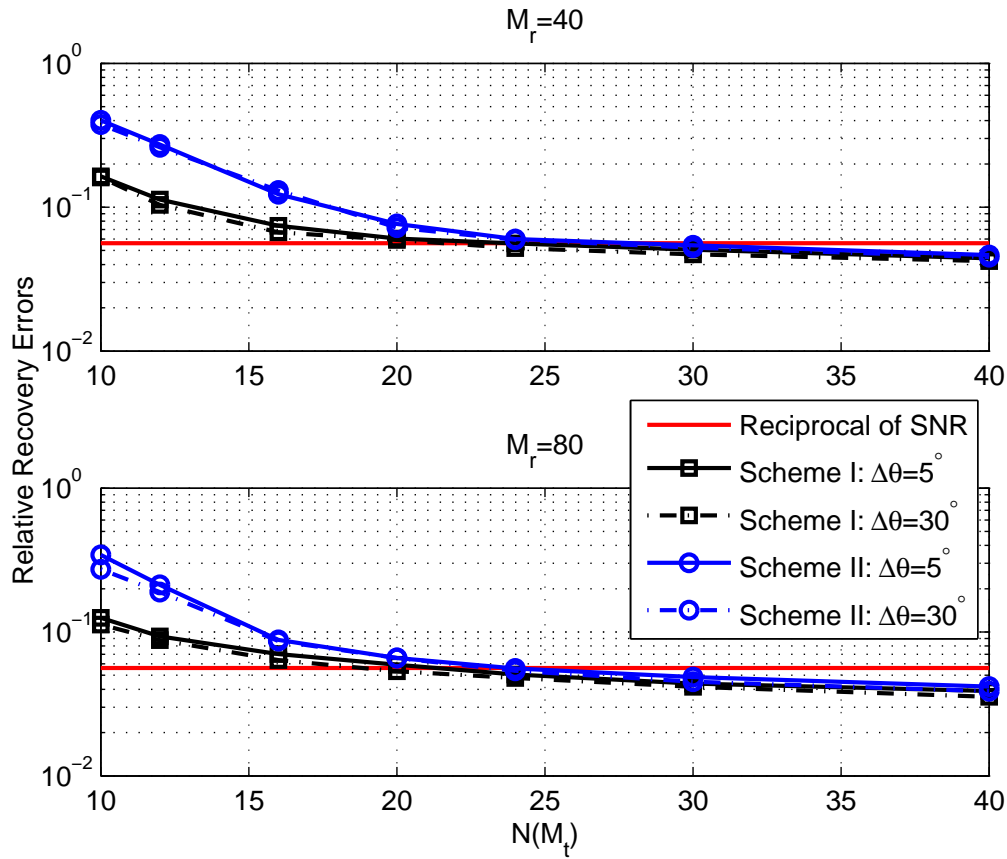}\caption{Comparions of the relative recovery errors
 in terms of number of $N$ ($M_t$) for $M_r=40,80$, respectively. The matrix occupy ratio is set as
 $p_1=p_2=0.5$.
  Two targets are generated at random in $\left[ { - 90^\circ,90^\circ} \right]$ with DOA separation $\Delta\theta=5^\circ,30^\circ$,
  respectively.} \label{error_comp}
\end{figure}

Comparing the two sampling methods based on the above figures (see Figs. \ref{DOA_resolution_I},
 and \ref{DOA_resolution_II} (a),(b))
 we see that although the performance is the same, sampling scheme I uses fewer samples,
 i.e.,  $10\times 20$ samples, as compared to sampling scheme II, which uses $128\times 20$ samples.
To further elaborate on this observation, we
 compare the performance  of the two sampling schemes when they both forward to the fusion center the same number of samples.
  The parameters are
   set  to $\rm{SNR}=25\rm{dB}$,  $p_1=p_2=0.5$ and  $M_t=N$. Therefore, in both schemes,
   the number of samples  forwarded by each receive antenna was the same.
   The number of transmit antenna was set as $M_r=40$ and $80$, respectively.
   Gaussian orthogonal transmit waveforms are
   used.
    Two targets are  generated at random in $\left[ { - 90^\circ,90^\circ} \right]$ at two different
   DOA separations, i.e.,  $\Delta\theta=5^\circ,30^\circ$.
   The results are averaged over $100$ independent realizations;
    in each realization, the targets are
   independently generated at random and the
   sub-sampling at each receive antenna is also independent between realizations.
    The relative recovery error comparison is plotted in Fig.\ref{error_comp}.

    It can be seen in Fig.\ref{error_comp} that as $N$ (or equivalently $M_t$) increases,
     the relative recovery error corresponding to $\tilde {\bf Z}_q$ and ${\bf Z}_q^{MF}$ decreases
     proportionally  to the reciprocal of the observed SNR.
     The relative recovery error under scheme I drops faster than  under scheme II for both
     $M_r=40$ and $M_r=80$ cases. This indicates that scheme
  I has a better performance than scheme II for the same number of samples.

\section{Conclusions}
\label{conclusion}

We have proposed MIMO-MC radars, which is a novel MIMO radar approach for high resolution target parameter estimation that involves small amounts of data.
Each
 receive antenna  either performs matched filtering with a small number of dictionary waveforms (scheme I) or
obtains sub-Nyquist samples of the received signal (scheme II) and forwards the results to a fusion center. Based on the samples forwarded by all receive nodes, and with
knowledge of the sampling scheme,
 the fusion center applies MC techniques  to estimate the full matrix, which can then be used in the context of existing array processing techniques, such as MUSIC, to obtain target information.
Although ULAs have been considered, the proposed ideas  can be generalized to arbitrary configurations. MIMO-MC radars are best suited for sensor networks with large numbers of
nodes.
Unlike MIMO-CS radars, there is no need for target space discretization, which avoids basis mismatch issues.
It has been confirmed with simulations that the coherence of the data matrix at the fusion center meets the conditions for MC techniques to be applicable. The coherence of the matrix
is always bounded by a small number. For scheme I, that number approaches $1$ as the number of transmit and receive antennas  increases and as the targets separation increases.
For scheme II, the coherence does not depend as much on the number of transmit and receive antennas,
or the target separation, but it does depend on $N$, the number of Nyquist samples within one pulse, which is related to the
bandwidth of the signal; the coherence increases as $N$ increases.
Comparing the two sampling schemes, scheme I has a better performance than scheme II for the same number of forwarded samples.

\appendix
\section{Proof of $\mu_0=\mu_1=1$ for a rank-$1$ ${\mathbf Z}_q^{MF}$}

\begin{proof}
Suppose that there are $K, K\geq 2$ targets in the search space, all  with the same DOA, say $\theta_1$.
The transmit and receive steering matrices are given by
\begin{align}
 {\bf{A}} &= \left[ {{\bf{a}}\left( {{\theta _1}} \right), \ldots ,{\bf{a}}\left( {{\theta _1}} \right)} \right], \\
 {\bf{B}} &= \left[ {{\bf{b}}\left( {{\theta _1}} \right), \ldots ,{\bf{b}}\left( {{\theta _1}} \right)} \right],
 \end{align}
where the transmit and receive steering vectors ${{\bf{a}}\left( {{\theta _1}} \right)}$ and ${{\bf{b}}\left( {{\theta _1}} \right)}$ are defined in equations (\ref{transmit_steering}) and (\ref{receive_steering}), respectively.
The noise-free receive data matrix equals
\begin{align}
  {\mathbf{Z}}_q^{MF} &= {\mathbf{B\Sigma }}{{\mathbf{D}}_q}{{\mathbf{A}}^T} \hfill \nonumber\\
   &= \left[ {{\bf{b}}\left( {{\theta _1}} \right), \ldots ,{\bf{b}}\left( {{\theta _1}} \right)} \right]\left[ {\begin{array}{*{20}{c}}
   {{\beta _1}} & {} & {}  \\
   {} &  \ddots  & {}  \\
   {} & {} & {{\beta _K}}  \\
\end{array} } \right]\left[ {\begin{array}{*{20}{c}}
   {{d_1}} & {} & {}  \\
   {} &  \ddots  & {}  \\
   {} & {} & {{d_K}}  \\
\end{array} } \right]\left[ {{\bf{a}}\left( {{\theta _1}} \right), \ldots ,{\bf{a}}\left( {{\theta _1}} \right)} \right]^T \hfill \nonumber\\
   &= \left( {\sum\limits_{k = 1}^K {{\beta _k}{d_k}} } \right){\bf{b}}\left( {{\theta _1}} \right){{\bf{a}}^T}\left( {{\theta _1}} \right), \hfill
\end{align}
where $d_k$ is the Doppler shift of the $k$-th target. Its compact SVD  is
\begin{align}
{\mathbf{Z}}_q^{MF} = {\mathbf{u}}\sigma {{\mathbf{v}}^H},
\end{align}
where ${{\mathbf{u}}^H}{\mathbf{u}} = 1,{{\mathbf{v}}^H}{\mathbf{v}} = 1$, and $\sigma$ is the singular value.

By applying the QR decomposition to the receive steering vector ${{\bf{b}}\left( {{\theta _1}} \right)}$, we have ${{\bf{b}}\left( {{\theta _1}} \right)} = {{\mathbf{q}}_r}{{{r}}_r}$, where ${\mathbf{q}}_r^H{{\mathbf{q}}_r} = 1$. Thus,
\begin{align}
{{\mathbf{q}}_r} = {\left[ {\frac{1}{{\sqrt {{M_r}} }},\frac{1}{{\sqrt {{M_r}} }}{\operatorname{e} ^{j\frac{{2\pi }}{\lambda }{d_r}\sin \left( {{\theta _1}} \right)}}, \ldots ,\frac{1}{{\sqrt {{M_r}} }}{\operatorname{e} ^{j\frac{{2\pi }}{\lambda }\left( {{M_r} - 1} \right){d_r}\sin \left( {{\theta _1}} \right)}}} \right]^T},
\end{align}
and ${r_r} = \sqrt {{M_r}} $.
Similarly, applying the QR decomposition to the transmit steering vector ${{\bf{a}}\left( {{\theta _1}} \right)}$, we have ${{\bf{a}}\left( {{\theta _1}} \right)} = {{\mathbf{q}}_t}{{{r}}_t}$, where ${\mathbf{q}}_t^H{{\mathbf{q}}_t} = 1$. Thus,
\begin{align}
{{\mathbf{q}}_t} = {\left[ {\frac{1}{{\sqrt {{M_t}} }},\frac{1}{{\sqrt {{M_t}} }}{\operatorname{e} ^{j\frac{{2\pi }}{\lambda }{d_t}\sin \left( {{\theta _1}} \right)}}, \ldots ,\frac{1}{{\sqrt {{M_t}} }}{\operatorname{e} ^{j\frac{{2\pi }}{\lambda }\left( {{M_t} - 1} \right){d_t}\sin \left( {{\theta _1}} \right)}}} \right]^T},
\end{align}
and ${r_t} = \sqrt {{M_t}} $.

Therefore, it holds that
\begin{align}
{\mathbf{Z}}_q^{MF} = {{\mathbf{q}}_r}\underbrace {{r_r}\left( {\sum\limits_{k = 1}^K {{\beta _k}{d_k}} } \right){r_t}}_\eta {\mathbf{q}}_t^T,
\end{align}
where $\eta$ is a complex number. Its SVD  can be written as $\eta  = {q_1}\rho q_2^*$, where $\left| {{q_1}} \right| = \left| {{q_2}} \right| = 1$, and $\rho$ is a real number. Thus,
\begin{align}
{\mathbf{Z}}_q^{MF} = {{\mathbf{q}}_r}{q_1}\rho q_2^*{\mathbf{q}}_t^T = {{\mathbf{q}}_r}{q_1}\rho {\left( {{\mathbf{q}}_t^*{q_2}} \right)^H},
\end{align}
where ${\left( {{{\mathbf{q}}_r}{q_1}} \right)^H}{{\mathbf{q}}_r}{q_1} = {\left| {{q_1}} \right|^2}{\mathbf{q}}_r^H{{\mathbf{q}}_r} = 1$ and ${\left( {{\mathbf{q}}_t^*{q_2}} \right)^H}{\mathbf{q}}_t^*{q_2} = {\left| {{q_2}} \right|^2}{\left( {{\mathbf{q}}_t^H{{\mathbf{q}}_t}} \right)^*} = 1$. By the uniqueness of the singular value, it holds that $\rho=\sigma$. Therefore, we can set ${\mathbf{u}} = {{\mathbf{q}}_r}{q_1}$ and ${\mathbf{v}} = {\mathbf{q}}_t^*{q_2}$.

Let ${{\mathbf{q}}_r^{\left( i \right)}}$ denote the $i$-th element of vector ${{\mathbf{q}}_r}$. The coherence $ \mu \left( U \right) $ is given by
\begin{align}
  \mu \left( U \right) &= \frac{{{M_r}}}{1}\mathop {\sup }\limits_{i \in {\mathbb{N}}_{{M_r}}^ + } \left\| {{\mathbf{q}}_r^{\left( i \right)}{q_1}} \right\|_2^2 \hfill \nonumber\\
   &= {M_r}\mathop {\sup }\limits_{i \in {\mathbb{N}}_{{M_r}}^ + } \left\| {{\mathbf{q}}_r^{\left( i \right)}} \right\|_2^2 \hfill \nonumber\\
   &= 1. \hfill
\end{align}
Let ${\mathbf{q}}_t^{*\left( i \right)}$ denote the $i$-th element of vector ${\mathbf{q}}_t^*$. The coherence $ \mu \left( V \right) $ is given by
\begin{align}
  \mu \left( V \right) &= \frac{{{M_t}}}{1}\mathop {\sup }\limits_{i \in {\mathbb{N}}_{{M_t}}^ + } \left\| {{\mathbf{q}}_t^{*\left( i \right)}{q_2}} \right\|_2^2 \hfill \nonumber\\
   &= {M_t}\mathop {\sup }\limits_{i \in {\mathbb{N}}_{M_t}^ + } \left\| {{\mathbf{q}}_t^{*\left( i \right)}} \right\|_2^2 \hfill \nonumber\\
   &= 1. \hfill
\end{align}
Consequently, we have ${\mu _0} = \max \left( {\mu \left( U \right),\mu \left( V \right)} \right) = 1$. In addition, we have ${\mu _1} \leq {\mu _0}\sqrt K  = 1$ \cite{Candes&Recht2009}. It always holds that $\mu_1\geq 1$. Thus, $\mu_1=1$.
Therefore, we have $\mu_0=\mu_1=1$.
\end{proof}

\vfill\break

%
%

%

%
%
%

\end{document}